\documentclass[11pt,a4paper,reqno]{amsart}

\usepackage{amsfonts,amsthm,amscd,epsfig,amsmath,amssymb,enumerate}

\numberwithin{equation}{section}

\DeclareMathSymbol{\leqslant}{\mathalpha}{AMSa}{"36} % nicer `smaller or equal'
\DeclareMathSymbol{\geqslant}{\mathalpha}{AMSa}{"3E} % nicer `larger or equal'
\DeclareMathSymbol{\eset}{\mathalpha}{AMSb}{"3F}     % nicer `emptyset'
\renewcommand{\leq}{\;\leqslant\;}                   % redef. of < or =
\renewcommand{\geq}{\;\geqslant\;}                   % redef. of > or =
\newcommand{\be}{\begin{equation}}

\def\1{\ifmmode {1\hskip -3pt \rm{I}} \else {\hbox {$1\hskip -3pt \rm{I}$}}\fi}

\newtheorem{Rem}{Remark}
\newcommand{\cA}{\ensuremath{\mathcal A}}

\newcommand{\cC}{\ensuremath{\mathcal C}}
\newcommand{\cD}{\ensuremath{\mathcal D}}

\newcommand{\cG}{\ensuremath{\mathcal G}}
\newcommand{\cH}{\ensuremath{\mathcal H}}

\newcommand{\cL}{\ensuremath{\mathcal L}}
\newcommand{\cM}{\ensuremath{\mathcal M}}

\newcommand{\cR}{\ensuremath{\mathcal R}}
\newcommand{\cS}{\ensuremath{\mathcal S}}
\newcommand{\cT}{\ensuremath{\mathcal T}}

\newcommand{\cX}{\ensuremath{\mathcal X}}
\newcommand{\cY}{\ensuremath{\mathcal Y}}

\newcommand{\bbE}{{\ensuremath{\mathbb E}} }
\newcommand{\bbF}{{\ensuremath{\mathbb F}} }

\newcommand{\bbP}{{\ensuremath{\mathbb P}} }

\newcommand{\bbR}{{\ensuremath{\mathbb R}} }

    \let\d=\delta  \let\e=\varepsilon
 \let\g=\gamma       \let\l=\lambda
      \let\o=\omega      
  \let\s=\sigma \let\t=\tau   
  
   \let\G=\Gamma   
\let\O=\Omega

\title[Piecewise deterministic Markov processes]
{Stationarity, time--reversal and fluctuation theory for a class of
piecewise deterministic Markov processes}

\author{A. Faggionato}
\address{Alessandra Faggionato. Dipartimento di Matematica ``G. Castelnuovo", Universit\`a ``La
  Sapienza''. P.le Aldo Moro  2, 00185  Roma, Italy. e--mail:
  faggiona@mat.uniroma1.it}
\author{D. Gabrielli}
\address{Davide Gabrielli.  Dipartimento di Matematica, Universit\`a
dell'Aquila,   67100 Coppito, L'Aquila, Italy. e--mail:
gabriell@univaq.it}
\author{M. Ribezzi Crivellari}
\address{Marco Ribezzi Crivellari. Dipartimento di Fisica,  Universit\`a Roma
Tre, Via della vasca navale 84, 00146 Roma . e--mail:
ribezzi@fis.uniroma3.it}

\begin{document}

\maketitle

\begin{abstract}
We consider a class of  stochastic dynamical systems, called
piecewise deterministic Markov processes,  with states  $(x, \s)\in
\O\times \G$,
 $\O$ being a region in $\bbR^d$ or the $d$--dimensional torus,
$\G$ being a finite set. The continuous variable $x$ follows a
piecewise deterministic dynamics, the discrete variable $\s$ evolves
by a stochastic jump dynamics  and the two resulting evolutions are
fully--coupled.  We study stationarity, reversibility and
time--reversal symmetries of the process. Increasing the frequency
of the $\s$--jumps, we show that the system behaves asymptotically
as deterministic and we investigate the structure of fluctuations
(i.e. deviations from the asymptotic behavior), recovering in a non
Markovian frame results obtained by Bertini et al. \cite{BDGJL1,
BDGJL2, BDGJL3, BDGJL4}, in the context of Markovian stochastic
interacting particle systems. Finally, we discuss a
Gallavotti--Cohen--type symmetry relation with involution map
different from time--reversal. For several examples the above
results are recovered by explicit computations.
\end{abstract}

\vskip0.5cm


\section{Introduction}
Piecewise deterministic Markov processes (PDMPs) are stochastic
dynamical systems whose  state is described by a pair $(x,\s)$,
where  $x$ is  a continuous variable and $\s$ is a discrete
variable. We take $x \in \O $ and $\s \in \G$, $\O$ being  a region
in $\bbR^d$ or the $d$--dimensional torus,  $\G$ being a finite set.
Motivated by applications to biochemical processes \cite{FGRbio}, we
call $x$ and $\s$ the mechanical and the chemical variable (or
state) of the system, respectively. The chemical state $\s$ evolves
by a random jump dynamics, while in the intervals in which the
chemical state is kept constant and equal to some $\s$, the
mechanical state $x$ evolves according to the deterministic
$\s$--dependent  ODE $\dot x (t)= F_\s (x(t))$. Since the
probability rates   of chemical jumps can depend on $x$, the
mechanical state $x$ and the chemical one $\s$ are dynamically
fully--coupled.  In our analysis, we restrict to time--homogeneous
PDMPs, i.e. both the vector fields $F_\s(x)$ and the probability
rate $\l r(\s,\s'|x)$ for a  chemical jump from $\s$ to $\s'$  at
the $x$--mechanical state  are time--independent. Above, $\l$ is a
positive parameter we will play with in order to analyze some
special regime. The above models can be used to describe the
overdamped motion of a particle in a viscous fluid under alternating
force fields, as well as some  biochemical processes as discussed in
\cite{FGRbio}

\smallskip

PDMPs are broadly used in applied sciences and engineering, and
%(see for example \cite{L}, \cite{N} and \cite{FGRbio}
% for applications to biochemical processes). In control theory, they
 are a typical example of what is called a {\it stochastic hydrid system} in control theory  \cite{CL}.
A mathematical analysis of PDMPs has been started by Davies in
\cite{D1} and the interested reader can find a detailed mathematical
treatment   in  \cite{D2}.  Our interest here is mainly  theoretical
and inspired by the physics of out--of--equilibrium systems. In
particular, our investigation concerns the steady state, the
time--reversed process, the deviations of the system from its
typical behavior, a special fluctuation--dissipation relation and a
Gallavotti--Cohen--type symmetry relation. We discuss our results in
more detail.

\smallskip

 In general, under mixing assumptions, the
steady state (stationary measure) is unique and has density $\rho
_\l (x,\s) $ on $\O \times \G$ which solves a system of differential
equations with zero--flux boundary conditions. We give exact
solutions   in dimension one, while for
 any  dimension we isolate a class of exactly solvable models for which
$\rho_\l$ has the special form
\begin{equation}\label{camaleonte} \rho_\lambda(x,\sigma)=c(\lambda)e^{-\lambda
S(x)}\rho(x,\sigma)\,, \end{equation}
 where $c(\lambda)$ is a
normalization factor depending only on $\lambda$.  Given the
stationary measure $\rho_\l$, we  consider the time--reversed
(adjoint) version of the PDMP and we  show  that it is again a PDMP,
with inverted vector fields and transition rates $\l r^+$ which
depend on the direct rates $r$ and on the stationary measure
$\rho_\l$. In particular, our  PDMPs typically  describe
out--of--equilibrium systems, since reversibility appears only in
the trivial case of vanishing force fields.  For exactly solvable
models as in \eqref{camaleonte}, we can show that
$r^+(\cdot,\cdot|x)$ does not depend on $\l$ as well some symmetry
relations.

\smallskip

In order to study the  deviations of the system from its relaxation
to equilibrium, we
  introduce a scaling procedure forcing the
system to behave deterministically in the asymptotic limit. Simply,
we  take the limit $\l\uparrow \infty$.
 As result,
 the timescale of chemical jumps becomes
infinitesimal w.r.t. the timescale of the mechanical evolution and
the  dynamics is a combination of slow and fast motions. As one
would expect, an {\it averaging principle} holds: the slow motion is
  well approximated by averaging the effect of the fast motion,
considering the fast (chemical) variable as locally equilibrated. In
order to be more precise, let us assume
  that for any $x\in \O$ the continuous--time
Markov chain on $\G$ with jump rates $r(\cdot,\cdot|x)$ ($x$ being
interpreted as frozen variable) is irreducible and therefore has a
unique invariant probability measure $\mu(\cdot |x)$ on $\G$. Then
the above high frequency limit implies that, whenever the mechanical
state of the PDMP is $x$,   the chemical state is given by  $\s$
with probability well approximated by $\mu(\s|x)$, while with
probability tending to $1$ the mechanical evolution $x(t)$ is well
approximated by  the deterministic path $x_*(\cdot)$ solving the
Cauchy system
\begin{equation}
\begin{cases}\label{detdet}
 \dot x_* (t)= \bar F ( x_* (t)) \,,\\
 x_*(0)=x_0\,,
 \end{cases}
 \end{equation}
 $x_0$ being the initial mechanical state and $\bar F$ being the
averaged vector field
$$ \bar F(x) = \sum_{\s \in \G} \mu(\s |x)F_\s (x)\,.
$$
 The above averaging principle corresponds  to    a law of large numbers
 for the mechanical evolution and, introducing suitable spaces and
topologies, it can be extended  to the joint evolution $(x(t),
\s(t))$ (see Section \ref{HFLLD}). A rigorous derivation of the
averaging principle as well as  the large deviations (LD) principle
for PDMPs can be found in the  companion paper \cite{FGRmat}.

\smallskip

 Considering the chemical variable as hidden and taking the limit $\l\uparrow\infty$, we  analyze the structure of
fluctuations of the mechanical variable, i.e.\ deviations from its
asymptotic deterministic behavior \eqref{detdet}, following ideas
and results of \cite{BDGJL1, BDGJL2, BDGJL3, BDGJL4}, for stochastic
interacting particle systems.
 A key identity observed in
\cite{BDGJL2} is  the  Fluctuation--Dissipation (FD) relation
\begin{equation}
\mathcal L(x,\dot{x})=\nabla V(x)\cdot \dot{x}+\mathcal
L^+(x,-\dot{x})\,,
\end{equation}
where $V$ denotes  the static LD functional of the steady state (in
exactly solvable models as in \eqref{camaleonte}, $V=S$), while
$\cL$ and $\cL^+$ are such that the LD functionals for the dynamics
of the PDMP and its time--reversed version are obtained by
integrating along the mechanical trajectories $x(t)$ the functions
$\cL(x,\dot{x})$ and $\cL^+(x, \dot{x})$, respectively. As observed
in \cite{BDGJL1, BDGJL2, BDGJL3, BDGJL4} and recalled in Section
\ref{nevissima}, whenever the FD relation is satisfied, several
physical properties concerning the relaxation of the system hold. In
\cite{BDGJL2} the authors derive the above FD relation from the
definition of the time--reversed process and  from the Markov
property of the processes under considerations (the direct one and
the time--reversed one). In our case, the mechanical evolution
$x(t)$ is typically non Markovian, hence the validity of the FD
relation has to be investigated.
 When
the dependence on the parameter $\l$ of the transition rates $\l
r^+$ is linear or almost linear, as in the case of 1D system or in
the case of exactly solvable model with stationary measure
\eqref{camaleonte}, we can apply again our  LD principle for
$\l$--rescaled PDMPs. Then,  we show the validity of the FD relation
    for the entire class of solvable
PDMPs whose stationary measures satisfies \eqref{camaleonte} as well
for PDMPs with two chemical states on the one dimensional torus for
which the stationary measure is not of the form \eqref{camaleonte}.

\smallskip

We briefly discuss a Gallavotti--Cohen--type symmetry relation for
PDMPs. The natural symmetry for this class of processes is different
from time-reversal and the corresponding action functional has a
direct physical interpretation. We obtain in this way examples
answering a question raised at the end of Subsection (2.2) in
\cite{LS}. See also \cite{M} and \cite{K} for more details and
references on the Gallavotti--Cohen symmetry in the framework of
stochastic dynamics.

\smallskip

The paper is structured as follows. In Section \ref{samarcanda} we
give a detailed description of the model and comment our basic
assumptions. In Section \ref{inv_e rev} we discuss stationarity and
reversibility, giving a system of differential equations with
boundary conditions characterizing the stationary measure and
showing that the adjoint process is again a PDMP. In Section
\ref{One} we compute the stationary measure for a class of 1D PDMPs,
which in many cases has the special form \eqref{camaleonte}. In
Section \ref{exactly} we outline a general method  valid in all
dimensions in order to determine stationary measures of the form
\eqref{camaleonte}, and we apply it in several examples. In Section
\ref{HFLLD} we recall the averaging and large deviation principles
obtained in \cite{FGRmat}. Having at this point all the necessary
tools, in Section \ref{nevissima} we study the statical and
dynamical deviations of the system from its typical behavior in the
same spirit of the fluctuation theory developed in \cite{BDGJL1,
BDGJL2, BDGJL3, BDGJL4}. Finally, in Section \ref{GC} we study a
Gallavotti--Cohen--type symmetry relation.

\smallskip

We conclude with a remark. Being stochastic dynamical systems, PDMPs
can have very different behaviors and show  special features. When
possible we have tried to keep our analysis at a general level, thus
requiring some mathematical abstraction. On the other hand,  special
dynamical mechanisms have been discussed directly by means of
examples. Moreover, even in very simple examples standard stochastic
tools used in the paper as  the Markov generator can become very
delicate and subtle. The interested reader can find some insights in
the appendix and can  refer to \cite{D2} for a general theory.

\section{The model}\label{samarcanda}
We consider stochastic models with state space $\Omega\times
\Gamma$, where $\G$ is a finite set and  $\Omega $ is either a
domain (i.e. open and connected subset) of $\mathbb R^d$ with
regular boundary $\partial \Omega$, or the closure of  a   domain of
$\mathbb R^d$ with regular boundary, or
  the $d$-dimensional torus $\mathbb
R^d/\mathbb Z^d$.
 A
generic element of the state space is denoted by $(x,\sigma)$.
Inspired by power--stroke models of molecular motors, we call  the
variables $x\in \Omega$ and $\sigma \in \G$  the   {\it mechanical
state} and the  {\it chemical state } of the system, respectively.
Their  joint stochastic evolution can be described as follows. The
mechanical state  $x$ evolves continuously, while the chemical state
$\s$ jumps at random times.
%The mechanical evolution is determined from the chemical one.
When the chemical state is $\sigma$,  the mechanical state evolves
according to the ordinary differential equation
\begin{equation}
\dot{x}(t)=F_{\sigma}(x(t)), \label{ode}
\end{equation}
where, for any $\sigma \in \Gamma$,
$F_{\sigma}(x)=\left(F_{\sigma,1}(x),\dots
,F_{\sigma,d}(x)\right)\in\mathbb R^d$ is a vector field. If $\O$ is
the $d$--dimensional torus, in the above equation $x$ is thought of
as element of a box in $\bbR^d$ with periodic boundary conditions.
We assume that the vector fields $F_\s$ have continuous extension to
the closure $\overline{\O}$ and satisfy the Lipschitz condition
\begin{equation}\label{lip}
|F_{\sigma}(x)-F_{\sigma}(y)|\leq K_{\sigma}|x-y|, \ \ \ \ \ \forall
x,y\in \Omega\,,
\end{equation}
for appropriate constants $K_{\sigma}$. Moreover,  we assume that
the mechanical evolution  remains confined  inside the region
$\Omega$. This assumption together with \eqref{lip} implies
existence and uniqueness of the mechanical trajectory.

\medskip

The chemical state $\s$  performs a jump stochastic dynamics with
rates depending on  the mechanical state. More precisely, the jump
rates are continuous functions  $r(\sigma,\sigma'|x):\Gamma\times
\Gamma\times \Omega \to \mathbb [0,\infty)$. Without loss of
generality we assume that
$$
r(\sigma,\sigma|x)=0\,, \ \ \ \ \forall (x,\sigma)\in \Omega\times
\Gamma\,.
$$
Moreover, we call  $\gamma(\sigma|x)=\sum_{\sigma'\in
\Gamma}r(\sigma,\sigma'|x)$. Given the initial state
$(x_0,\sigma_0)\in \O\times \G$, we consider the  random variable
$\tau_1$ with distribution
\begin{equation}
\mathbb P(\tau_1> t)=\left\{
\begin{array}{lc}
e^{-\lambda\int_0^t\gamma(\sigma_0|x_0(s))ds} & t\geq 0\,,\\
1 & t<0\,.\\
\end{array}
\right.\nonumber
\end{equation}
In the above formula $\lambda$ is a positive parameter  and $x_0(s)$
is the solution of the Cauchy problem
\begin{equation}
\left\{
\begin{array}{l}
\dot{x}(t)= F_{\sigma_0}(x(t))\,, \\
x(0)=x_0 \,.
\end{array}
\right.\label{Cauchy}
\end{equation}
 The evolution of the system (mechanical state and chemical
state) in the time interval $[0,\tau_1)$ is given by
$(x_0(s),\sigma_0)$. The chemical state $\sigma(\tau_1)$ is then
chosen in $\Gamma$ according to the distribution
$$
\mathbb
P(\sigma(\tau_1)=\sigma)=\frac{r(\sigma_0,\sigma|x(\tau_1))}{\gamma(\sigma_0|x(\tau_1))}\,.
$$
Let $x_1(t)$, $t \geq \t_1$, be the solution of \eqref{Cauchy} for
the vector field $F_{\sigma(\tau_1)}$ and initial condition
$x_1(\tau_1)=x_0(\tau_1)$. Let $\tau_2$ be a random variable
distributed according to
\begin{equation}
\mathbb P(\tau_2>
t)=\left\{
\begin{array}{lc}
e^{-\lambda\int_{\tau_1}^{t}\gamma(\sigma(\tau_1)|x_1(s))ds} & t\geq
\tau_1\,,\\
1 & t<\tau_1\,.\\
\end{array}
\right.\nonumber
\end{equation}
The evolution of the system in the time interval $[\tau_1,\tau_2)$
is given by $(x_1(s),\sigma(\tau_1))$. The chemical state
$\sigma(\tau_2)$ is chosen in $\Gamma$ according to the distribution
$$
\mathbb
P(\sigma(\tau_2)=\sigma)=\frac{r(\sigma(\tau_1),\sigma|x(\tau_2))}{\gamma(\sigma(\tau_1)|x(\tau_2))}\,,
$$
and so on. In general, we denote by $\tau_k$ the random time of the
$k$--th chemical jump and by $(x_k(s), \sigma(\tau_{k}))$ the state
of the system at time $s\in [\tau_k,\tau_{k+1})$. We have that
$x_k(s)$ solves \eqref{Cauchy} for the vector field
$F_{\sigma(\tau_k)}$ with initial condition
$x_k(\tau_k):=x_{k-1}(\tau_k)$ and that $\tau_{k+1}$ is a random
variable with distribution
\begin{equation}\label{sanna}
\mathbb P(\tau_{k+1}> t)=
\begin{cases}
e^{-\lambda\int_{\tau_k}^{t}\gamma(\sigma(\tau_k)|x_k(s))ds} & t\geq
\tau_k\,,\\
1 & t<\tau_k\,.
\end{cases}
\end{equation}
 The chemical state $\sigma(\tau_{k+1})$ is then chosen in $\Gamma$
according to the distribution
$$
\mathbb
P(\sigma(\tau_{k+1})=\sigma)=\frac{r(\sigma(\tau_{k}),\sigma|x(\tau_{k+1}))}{\gamma(\sigma(\tau_k)|x(\tau_{k+1}))}\,.
$$
In order to have a well--defined dynamics for all positive times we
require that a.s. the family of jump times $\t_k$ has no
accumulation point.  This is always true if $\lim_{k\uparrow
\infty}\tau_k=+\infty$ a.s.

\medskip

The natural path space of the process $\bigl( x(\cdot), \s(\cdot)
\bigr)$ is given by the cartesian product
\begin{equation}
C([0,T],\Omega)\times D([0,T], \Gamma)\,. \label{cart}
\end{equation}
The first component is the space of continuous functions from
$[0,T]$ to $\Omega$, while the second component is the Skorokhod
space of functions from $[0,T]$ to $\G$, which are   right
continuous and have  left limits.
% A trajectory of a PDMP is
%$\left\{(x(t),\sigma(t))\right\}_{t\in[0,T]}$ where
%$\left\{x(t)\right\}_{t\in[0,T]}\in C([0,T],\mathbb R^d)$ and
%$\left\{\sigma(t)\right\}_{t\in[0,T]}\in D([0,T], \Gamma)$. In fact,
We note that, due to relation \eqref{ode}, the mechanical trajectory
$\left\{x(t)\right\}_{t\in[0,T]}$ is a piecewise differentiable
function and it holds
\begin{equation}
\dot{x}(s)=F_{\sigma(s)}(x(s))\,, \label{ad}
\end{equation}
for any $s\in [0,T]$ where $\left\{\sigma(t)\right\}_{t\in[0,T]}$ is
continuous.

\medskip

The above stochastic  process $\bigl(x(\cdot), \s (\cdot) \bigr)$ is
called {\it Piecewise Deterministic Markov Process} (PDMP)
\cite{D1}, \cite{D2}. In control theory, it is a typical example of
{\it stochastic hybrid system} \cite{CL}. Although the evolution of
the mechanical state as well the evolution of the chemical state are
not Markovian,
 as proven in \cite{D1}, \cite{D2} the joint evolution $\bigl(x(\cdot), \s (\cdot) \bigr)$  is a strong Markov process, whose
  Markov generator is formally
\begin{equation}
Lf(x,\sigma)=F_\sigma(x)\cdot \nabla
f(x,\sigma)+\lambda\sum_{\sigma'\in
\Gamma}r(\sigma,\sigma'|x)\left(f(x,\sigma')-f(x,\sigma)\right)
\label{gen}
\end{equation}
where $f:\Omega\times \Gamma\to \mathbb R$ is regular  in the
mechanical  variable  $x$. In Appendix \ref{ext_gen} we will specify
the precise meaning and domain of the operator $L$.
% A characterization
%of the domain of the so called extended generator of the process is
%given in \cite{D1}, \cite{D2}.
For any fixed $x\in \Omega$,  the $x$-dependent Markov generator
$L_c[x]$ on $\Gamma$ given by
\begin{equation}
L_c[x]f(\sigma)=\sum_{\sigma'\in
\Gamma}r(\sigma,\sigma'|x)\left(f(\sigma')-f(\sigma)\right)
\label{genchem}
\end{equation}
is called the {\it chemical part} of the generator $L$. Hence, we
can write
$$
Lf(x,\sigma)=F_\sigma(x)\cdot \nabla f(x,\sigma)+\lambda
L_c[x]f(x,\sigma).
$$
Note that $L_c[x]$ is the Markov generator of a continuous--time
Markov chain on $\G$ where jumps from $\s$ to $\s'$ take place with
probability rate $r(\s,\s'|x)$. We will assume that for any fixed
$x\in \Omega$, this Markov chain
 is irreducible and consequently has a unique stationary measure
$\mu(\cdot|x)$,  that we call   {\it quasistationary measure}. In
the  particular case of two chemical states, e.g. $\G=\{0,1\}$, this
condition reduces to the positivity of the rates $r(0,1|x)$ and
$r(1,0|x)$. In this case, the quasistationary measure is also
reversible w.r.t. the Markov chain on $\G$ with generator $L_c[x]$
and is given by \begin{equation}\label{rev1D}
 \mu(0|x) = \frac{
r(1,0|x)}{r(0,1|x)+r(1,0|x)}\,, \qquad \mu(0|x) = \frac{
r(0,1|x)}{r(0,1|x)+r(1,0|x)}\,.
\end{equation}

\medskip

Let us now come back to our assumptions and give some comments. As
already stated, in order to have a well--defined dynamics for all
positive times we
 require  that  a.s.  the family of random jump times $\t_k$ has no accumulation point.
 This fact is implied
 for example by the condition
 \begin{equation}\label{sciopero0}
\sup _{\{x\in \O\}}\max_{\{\s \in \G\}} \g (\s|x) <\infty\,.
\end{equation}
Indeed, calling $C$ the l.h.s. of \eqref{sciopero0}, due to
 \eqref{sanna} we get that  $\bbP ( \t_{k+1}-\t_k >t ) \geq e^{-C\l
 t}$. This allows to build a coupling between the  family of random
 jump times $\t_k$   and
  a Poisson point process (PPP) on $(0,\infty)$ with density $C\l $ such
 that all jump times $\t_k$ belong to the PPP.
 Since a.s. the PPP has no accumulation point,
 the same property holds for the family of jump times $\t_k$ and
 this proves our claim. Let us also point out that, due to the above coupling, we get that the random variable
 $N_t$ given  by the number of jump times $\t_k$ in the time interval $[0,t]$
 has finite expectation.

One can even weaken condition \eqref{sciopero0}.
 Due to the irreducibility assumption for
$L_c [x]$, the absence of accumulation points for the family of jump
times $\t_k$  is implied by the condition
\begin{equation}\label{sciopero}
 \sup_{\left\{x\in
\Omega\right\}}\min_{\left\{\sigma \in
\Gamma\right\}}\gamma(\sigma|x)<+\infty\end{equation} and some other
additional assumptions. Without trying to give some general
criterion, in order to explain the mechanism we have in mind  we
discuss in Appendix \ref{papi}   an example where \eqref{sciopero}
is valid and \eqref{sciopero0} is violated, while the family of jump
times $\t_k$ has no accumulation point.

As already stated, we assume that the mechanical trajectory $x(t)$
remains confined inside $\O$. If $\O$ is the $d$--dimensional torus,
this assumption is trivially satisfied. Let us consider  the case
$\O\subset \bbR^d$.
Then, it is necessary that there is zero flux through the boundary.
A sufficient condition is given by
\begin{equation}
F_{\sigma}(x)\cdot n(x) \leq 0, \ \ \ \forall \sigma\in \Gamma, \; x
\in \partial \Omega\,, \label{scabound}
\end{equation}
where $\cdot$ denotes the Euclidean scalar product in $\mathbb R^d$,
while  $n(x)$  denotes the outward normal to $\partial \Omega$. If
$\O$ includes its boundary $\partial \O$, the above condition is
enough to have confinement, otherwise one can  require in addition
to \eqref{scabound}  that
\begin{equation}\label{cipcip}
F_\s (x) =0 \qquad \forall x\in \partial\O\,\text{ s.t. }  F_\s (x)
\cdot n(x) =0 \,. \end{equation} This condition excludes the
presence of orbits tangent to the boundary at some point.

Another condition assuring   the confinement of $x(t)$ is the
following. We take $\O\subset \bbR^d$ open for simplicity. For any
$\sigma\in \G$ and $x_0\in
\partial \Omega$, consider the trajectory $x(t)$ starting in $x_0$
with reversed vector field $-F_\s(x)$, namely $\dot x (t)=-F_\s
(x(t))$. If this trajectory is well defined in some  time interval
$[0,t_0)$ such that  $x(t)\in \O$ for all $t\in (0,t_0)$, then we
require that
\begin{equation}
\int_0^{t_0} \gamma(\sigma|x(s))ds =+\infty\,. \label{nonlomesso}
\end{equation}
We claim that  the mechanical evolution is kept bounded inside
$\Omega$ by a stochastic mechanism. In fact, condition
\eqref{nonlomesso} guarantees that if the mechanical trajectory is
pointing towards $x_0$,   with probability one there is a  jump to a
new chemical state before reaching $x_0$. Now again, if the
mechanical trajectory associated to the  new chemical state  is
pointing towards a point $x_1\in \partial \O$, with probability one
there is a jump to a new chemical state before reaching $x_1$. In
order to reach the boundary $\partial \O$ in a finite time, the
system should perform infinite chemical jumps in that time interval,
which is not possible due to our assumptions.
In the examples discussed below,   the above two criteria for the
mechanical  confinement inside open $\O$'s are dual: if the system
is confined due to \eqref{scabound} and \eqref{cipcip}, the
time--reversed system will be confined due to \eqref{nonlomesso}.

\medskip

We will mainly be interested in models such that a.s. the family of
random jumps times $\tau_k$ is infinite. A sufficient condition to
obtain this behavior is given by
\begin{equation}
\inf_{\left\{x\in \O\right\}}\min_{\left\{\sigma\in
\Gamma\right\}}\gamma(\sigma|x)>0\,. \label{bonbon}
\end{equation}
In fact, with arguments similar to the ones after \eqref{sciopero0},
it can be shown that the family of jump times dominates a PPP. We
discuss in Appendix \ref{apeape} an example of a PDMP that violates
\eqref{bonbon} and has a.s. a finite number of chemical jumps.

\medskip

Finally we remark that if there exists   $x^*\in \O$ such that
$F_\s(x^*)=0$ for any $\s\in \Gamma$, then the PDMP with initial
condition $x(0)=x^*$ has a trivial behavior: the mechanical variable
does not evolve, while  the chemical variable  evolves as a
continuous--time Markov chain with transition rates
$r(\sigma,\sigma'|x^*)$.

\medskip

We conclude this section with some notation frequently used below.
Given a point $(x,\s)\in \O \times \G$, we denote by $\bbP^\l_{x,\s}
$ and $\bbE ^\l _{x,\s}$ the law of the process $\bigl( x(\cdot), \s
(\cdot) \bigr)$ starting in $(x,\s)$ and the associated expectation,
respectively.

\section{Stationarity  and reversibility}
\label{inv_e rev}

A probability measure $\rho_\lambda$ on $\Omega\times \Gamma$ is
called {\sl invariant}, or {\sl stationary},  if  for any time $t
\geq 0$ the pair $(x(t),\sigma(t))$ is distributed according to
$\rho_\lambda$ when the process has initial distribution
$\rho_\lambda$. In this case, the process starting with distribution
$\rho_\lambda$ is called stationary.
Since the dynamics is defined by
time--independent rules, this implies that for each $t\geq 0$ the
laws of the  trajectory  $\bigl\{ x(s), \s(s) \bigr\}_{s\geq 0}$ and
the law of the time--shifted trajectory $\bigl\{ x(s+t), \s (s+t)
\bigr\}_{s \geq 0}$ coincide. This observation together with
Kolmogorov Theorem  allows to univocally extend  the process for all
times $t \in \bbR$, by requiring that its law  does not change under
time--shifts. Below, when referring to a stationary process we will
often mean the $\bbR$--extended one. In particular, if the process
is stationary we can define its time--reversed version. To this aim,
we introduce the {\it time--reversal operator}, defined as
\begin{equation}
\mathcal T \bigl[\bigl\{x(t),\sigma (t)\bigr\}_{t\in \bbR
}\bigr]:=\bigl\{x(-t),\sigma (-t)\bigr\}_{t\in \bbR
}\,.\label{timer}
\end{equation}
Since we want  $\mathcal T$ to be an operator from and onto the path
space $C( \bbR, \O) \times D( \bbR, \G)$, the above definition has
to be slightly modified by replacing the r.h.s. of \eqref{timer}
with the only path in  $C( \bbR, \O) \times D( \bbR, \G)$ coinciding
with $\bigl\{x(-t),\sigma (-t)\bigr\}_{t\in \bbR }$ at all
continuity points $t$. For the sake of simplicity, when writing the
r.h.s. of \eqref{timer} we will mean this last path. Then the {\sl
time--reversed process} (also called {\sl adjoint process}) is
defined as the process obtained from the original one via
  $\cT$. If $\bbP^\l_{\rho_\l}$ is the law of the original process,
  then $\bbP^{\l,+}_{\rho_\l}:=\bbP^\l_{\rho_\l}\circ \cT^{-1}$ is the law of the
  time--reversed process. If the two laws coincide, one says  that the
  process with law $\bbP^\l_{\rho_\l}$ is {\it reversible} and that
  $\rho_\l$ is a reversible measure for the process. As we will
  explain below, the time--reversed process is again a PDMP. Due to
  the Markov property, by considering the transition probability
  kernels of $\bbP_{\rho_\l} ^{\l,+}$, one can easily define the adjoint process with arbitrary
  initial distribution $\nu$. We will denote $\bbP^{\l,+}_\nu$ its
  law on  $C( \bbR_+, \O) \times D( \bbR_+, \G)$.

\smallskip

 Existence and uniqueness of the stationary measure of a PDMP can be
 non
 trivial and the analysis can be very model--dependent. If $\O$ is
 bounded one can deduce the existence of a stationary measure as follows.
Fixed any initial distribution $\nu$, we write $\nu_s$ for the law
of $(x(s),\s(s))$ under $\bbP^\l _\nu$. By compactness arguments
(cf. \cite{Bil}),  the family of Cesaro averages $ \tilde
{\nu}_t:=t^{-1} \int _0^t \nu_s ds $ admits a subsequence
$\tilde{\nu}_{t_n}$, with $t _n \uparrow \infty$, converging to some
probability measure $\nu_*$ on $\bar \O \times \G$ . If  $\O$ is
closed  or if $\O$ is open and $\nu_* (\partial \O)=0$, then by
standard arguments one can show that $\nu_*$ is a stationary
probability measure for the PDMP. The fact that
$\nu_*(\partial\O)=0$ for $\O\subset \bbR^d $ can be verified for
example if, when  approaching the boundary, the system typically
jumps to a chemical state $\s$ with a vector field $F_\s$ of order
one pointing inside $\O$. We will come back to this mechanism in
Appendix \ref{papi}. Other existence criteria are given in
\cite{D2}.

\smallskip

In the rest of this section we  restrict to the case of $\O\subset
\bbR^d $ open or $\O$ being a $d$--dimensional torus. Assuming now
the existence of some stationary measure $\rho_\l$, we want to give
a characterization of $\rho_\l$ by means of partial differential
equations.  Due to Theorem 34.19 in \cite{D2}, if $\rho_\l$ is a
stationary measure then it must be
\begin{equation}\label{uuu}
 \rho_\l (L f) =
0 \end{equation}
 for  a large class of functions $f$ in the
domain of the extended  generator (see Appendix \ref{ext_gen} for
the definition of extended generator) and  in particular for
functions  $f$ which are  bounded and $C^1$ in $x$. Let us denote
here by $\bbF$ the family of functions $f$ which are $C^1$ in $x$,
have continuous extension to $\partial \O$ and satisfy for some $\s$
the property: $f(x,\s')=0$ if $\s'\not = \s$. Assuming $ \rho_\l =
\sum _\s \rho _\l (x,\s) dx \d_{\s}$ (i.e. $\rho_\l$ is absolutely
continuous w.r.t. the Lebesgue measure on each subspace $\O\times
\{\s\}$) and assuming that $\rho_\l (\cdot, \s)$ is $C^1$, equation
\eqref{uuu} for $f \in \bbF$ reads
\begin{equation}\label{lattino}
\int _\O dx\, \rho_\l (x,\s) F_\s (x) \cdot \nabla f (x,\s) +\l
\sum_{\s'\in \G} \int _\O dx f(x,\s) \bigl( \rho_\l
(x,\s')r(\s',\s)- \rho_\l (x,\s) r(\s,\s') \bigr) =0\,.
\end{equation}
By the Gauss--Green formula, the first integral in the l.h.s. equals
\begin{equation}\label{GG}
\int_{\partial \O} dS(x)   f(x,\s) \rho _\l (x,\s) F_\s (x) \cdot
n(x) - \int _\O dx f(x,\s) \nabla \cdot \bigl( \rho_\l (x,\s) F_\s
(x) \bigr)\,,
\end{equation}
where $dS$ denotes the $(d-1)$--dimensional surface measure on
$\partial \O$ and $n(x)$ denotes the outward normal to $\partial \O$
in $x$. Since \eqref{lattino} must hold in particular for all
functions $f\in \bbF $ with $x$--support given by a compact subset
of $\O$, we conclude that
\begin{equation}
\lambda \sum_{\sigma'\in
\Gamma}\Big(\rho_\lambda(x,\sigma')r(\sigma',\sigma|x)-\rho_\lambda(x,\sigma)r(\sigma,\sigma'|x)\Big)=\nabla
\cdot\Big(\rho_\lambda(x,\sigma)F_{\sigma}(x)\Big), \ \ \ \ \forall
(x,\sigma)\in \Omega\times \Gamma. \label{inv}
\end{equation}
Then, due to \eqref{lattino}, \eqref{GG} and \eqref{inv}, the
boundary integral in \eqref{GG} must be zero for all functions $f
\in \bbF$. This forces $\rho_\l$ to have   zero flux across the
boundary $\partial \O$:
\begin{equation}
\Big(\rho_\lambda(x,\sigma)F_{\sigma}(x)\cdot n(x)\Big)\Big|_{x\in
\partial \O}=0, \ \ \ \ \ \forall \sigma\in \Gamma.
\label{bondcond}
\end{equation}
Note that if $\O$ is the closure of a domain in $\bbR^d$, the system
of identities \eqref{inv} must still be valid for $(x,\s)\in
\Omega^\circ\times \Gamma$, $\Omega^\circ$ being the interior part
of $\Omega$, since it follows from \eqref{uuu} by taking arbitrary
functions $f$ which are bounded, $C^1$ in $x$ and with $x$--support
strictly included in $\O^\circ$. The boundary condition can differ
from \eqref{bondcond}, depending on the dynamics at the boundary.

\smallskip

As already mentioned, if the process $\bbP^\l_{\rho_\l} $ is
stationary, then its time--reversed version is again a PDMP. More
precisely,  we claim that $\bbP^{\l, +}_{\rho_\l}$ coincides with
the law $P$ of the  PDMP with state space $\O \times \G$, initial
distribution $\rho_\l$,  force fields
\begin{equation}\label{forza+}
 F^+_\s
(x):=-F_\s(x)\end{equation} and jump rates $\l r^+ (\s,\s'|x)$ where
\begin{equation}
r^+(\sigma,\sigma'|x):=r(\sigma',\sigma|x)\frac{\rho_\lambda(x,\sigma')}{\rho_\lambda(x,\sigma)}\,.
\label{defr+}
\end{equation}
(In the above formula, we assume some spatial irreducibility of the
system implying the positivity of  $\rho_\l (x,\s)$).
 Note that writing
\begin{equation}\label{defg+} \g^{+}(\s|x):=\sum
_{\s'}r^+(\sigma,\sigma'|x)\,,
\end{equation}
equation \eqref{inv} can be written as
\begin{equation}\label{invbis}
\l \Big(\g^+ (\s|x)- \g(\s|x)\Big)= \frac{ \nabla \cdot \bigl(
\rho_\l (x,\s) F_\s (x) \bigr)}{ \rho_\l (x,\s) }\,.
\end{equation}

In order to prove our claim, due to the stationarity of
$\bbP^\l_{\rho_\l}$  we only need to show that
\begin{equation}\label{siriano}
 P (A)=\bbP^\l_{\rho_\l} (B) \,,
\end{equation}
where the events $A$ and $B$ are of the following form. Fix a
measurable set $U\subset \O$, $k+1$ chemical states $\s_0, \s_1,
\dots, \s_{k}$ with $\s_i \not = \s_{i+1}$, $k$ positive times
$0<t_1<\cdots < t_k<T$ and $\epsilon$ small enough. Define $A$ as
the family of paths $(x(\cdot), \s(\cdot) )\in C(\bbR_+, \O) \times
D(\bbR_+, \G)$ such that (i) there are exactly $k$ jump times $\t_1,
\t_2, \dots ,\t_{k}$ in the time interval $(0,T)$ and $\t_i\in
[t_i-\epsilon, t_i+\epsilon]$, (ii) $\s(\t_i)=\s_i$ for $0< i \leq
k$ and moreover $\s(0)=\s_0$, (iii) $x(0)\in U$ and (iv) $\dot{x}
(s) = -F_{\s(s) } (x(s))$ for all $s \in [0,T)$ different from the
jump times. Then define $B$ as the family of paths $(x(\cdot),
\s(\cdot) )$ such that (i) there are exactly $k$ jumps times $\t'_1,
\dots, \t'_k$ in the time interval $[0,T]$ and $\t'_i\in
[T-t_{k-i+1}-\epsilon,T-t_{k-i+1}+\epsilon]$, (ii) $ \s(\t'_i
)=\s_{k-i}$ for $1\leq i \leq k$ and moreover $\s(0)=\s_k$, (iii)
$x(T)\in U $ and (iv) $\dot{x} (s) = F_{\s(s) } (x(s))$ for all $s
\in [0,T]$ different from the jump times. Trivially, apart
time--shift $A$ and $\cT ^{-1} B$ coincide. Note that the validity
of \eqref{siriano} for all events $A$ and $B$ as above implies also
that $P$ gives probability $1$ to paths with mechanical variable
confined in $\O$.

\smallskip

For the  sake of simplicity and without loss of generality, in order
to prove \eqref{siriano} we take $k=1$. Moreover, given $x_0 \in U$
and $t\in [t_1-\epsilon,t_1+\epsilon]$, we write $\bigl(x(\cdot),
\s(\cdot) \bigr)$ for the unique path in $A$ jumping at time $t$ and
such that $x(0)=x_0$. Then, due to \eqref{defr+}, \eqref{defg+} and
\eqref{invbis}, we conclude that
\begin{equation}
P(A)=\int_{t_1-\epsilon}^{t_1+\epsilon}g_{t} dt\,, \label{dede}
\end{equation}
where the density $g_t$ is given by
\begin{equation}\label{tosse}
\begin{split} g_t & =\l  \int _U dx_0 \rho_\l (x_0, \s_0)
r^+ \bigl(\s_0,\s_1 | x(t) \bigr)\exp\bigl\{ -\l \int _0 ^{T} \g^{+}
(\s(s)
| x(s) ) ds \bigr\}\\
& =\l
 \int _U dx_0 \rho_\l (x_0, \s_0)\frac{\rho_\l ( x(t), \s_1)
}{\rho_\l (x(t), \s_0) }  r \bigl(\s_1,\s_0 | x(t) \bigr)
\qquad \qquad \qquad \qquad \qquad \qquad  \\
& \times \exp\Big\{ -\l \int _0 ^{T} \g (\s(s) | x(s) ) ds \Big \}
\exp \Big\{  - \int_0^{T} \frac{ \nabla\cdot \Big( \rho_\l
\bigl(x(s),\s(s)\bigr ) F_{\s(s) }(x(s)) \Big) }{\rho_\l (x(s),
\s(s) ) }ds \Big\} \,.
\end{split}
\end{equation}
We note that in the time intervals $(0,t)$ and $(t,T)$ where the
chemical state is constant, it holds
$$
\frac{ \nabla\cdot \Big( \rho_\l \bigl(x(s),\s \bigr ) F_{\s }(x(s))
\Big) }{\rho_\l (x(s), \s  ) }=- \frac{d}{ds} \ln \rho_\l \bigl(
x(s), \s \bigr)+ \nabla\cdot F_\s (x(s))\,,
$$
where $\s$ is the value of the chemical state.
 Dividing the interval
of integration $[0,T]$ in the two intervals $[0,t]$ and $[t,T]$, the
above identity implies   that the last exponential factor in
\eqref{tosse} equals
$$
\frac{ \rho_\l (x(t) , \s_0) \rho_\l (x(T), \s_1) }{ \rho_\l (x_0,
\s_0) \rho_\l ( x(t) , \s_1)  }\exp\Big\{ - \int_0^{T} \nabla \cdot
F_{\s(s)} (x(s)) ds \Big\}\,.
$$
This observation together with \eqref{tosse}  allows to conclude
that
\begin{multline}\label{fanalino}
 g_t = \l \int _U dx_0 \rho_\l (x(T), \s_1) r \bigl(\s_1,\s_0 | x(t)
 \bigr)\\\times
 \exp\bigl\{ -\l \int _0 ^{T} \g (\s(s) | x(s) ) ds \bigr
\} \exp \bigl\{- \int_0^{T} \nabla \cdot F_{\s(s)} (x(s)) ds
\bigr\}\,.
\end{multline}
Let us now consider the diffeomorphism $\phi :U\to \phi(U)$ that
associate to $x_0\in U$ the element $x(T)\in \phi (U)$. This is
obtained from the composition $\phi_2 \circ \phi_1$ where $\phi_1$
maps $x_0$ to $x(t)$ and $\phi_2$ maps $x(t)$ to $x(T)$. By
Liouville theorem  the Jacobian $J_{\phi_1}$ of the diffeomorphism
$\phi_1$ is given by
\begin{equation}
J_{\phi_1}(x_0)=e^{-\int_0^t\nabla \cdot F_{\s_0}(x(s)) ds}\,,
\end{equation}
and the Jacobian $J_{\phi_2}$ of the diffeomorphism $\phi_2$ is
given by
\begin{equation}
J_{\phi_1}(x(t))=e^{-\int_t^T\nabla \cdot F_{\s_1}(x(s)) ds}\,.
\end{equation}
If we perform in \eqref{fanalino} the change of variables
$y_0=\phi(x_0)$ we obtain
\begin{equation}\label{fanalinobis}
 g_t =\l  \int _{\phi(U)} dy_0 \rho_\l (y_0, \s_1) r \bigl(\s_1,\s_0 | y(T-t)
 \bigr)
 \exp\bigl\{ -\l \int _0 ^{T} \g (\s(s) | y(s) ) ds \bigr
\}:=\widetilde{g}_{T-t}\,,
\end{equation}
where $\left(y(\cdot),\s(\cdot)\right)$ is the unique path in $B$
jumping at time $T-t$ and such that $y(0)=y_0$. Now \eqref{siriano}
follows from \eqref{dede} and the identity
$$
\mathbb
P^\l_{\rho_\l}(B)=\int_{T-t_1-\epsilon}^{T-t_1+\epsilon}\widetilde{g}_t
dt\,.
$$

\medskip

A different route in order to characterize the time--reversed
process is given by the analysis of the generator. Being aware   of
the subtle difficulties concerning the domain of definition of the
generator (see Appendix \ref{ext_gen}), we keep this analysis at a
very heuristic level. The generator $L^+$ of the time--reversed
process must be the adjoint  in $L^2 (\rho_\lambda)$ of the
generator $L$ of the direct process,
  namely
\begin{equation}
\mathbb E_{\rho_\lambda}\left(gLf\right)=\mathbb
E_{\rho_\lambda}\left(fL^+g\right)\label{adgio}
\end{equation}
for all $f,g$ regular enough. Below we  check that this implies the
identity
\begin{equation}
L^+f(x,\sigma)=-F_{\sigma}(x)\cdot \nabla
f(x,\sigma)+\lambda\sum_{\sigma'\in
\Gamma}r(\sigma',\sigma|x)\frac{\rho_\lambda(x,\sigma')}{\rho_\lambda(x,\sigma)}\Big(f(x,\sigma')-
f(x,\sigma)\Big). \label{adgiogen}
\end{equation}
We can write \eqref{adgiogen} as
\begin{equation}
L^+f(x,\sigma)=-F_\sigma(x)\cdot \nabla f(x,\sigma)+\lambda
L_c^+[x]f(x,\sigma)\,. \label{adgiogen2}
\end{equation}
We stress  that $L_c^+[x]$ denotes  the chemical part of the adjoint
generator $L^+$. In general, this  is different from  the operator
$L_c[x]^+$, defined as the adjoint in $L^2 (\mu (\cdot |x) )$ of the
chemical part $L_c[x]$ of the generator $L$:
\begin{equation}\label{attenzione}
L_c ^+[x]\not= L_c [x]^+\,.
\end{equation}

The general formula \eqref{adgiogen} can be easily checked as
follows. We start from the left hand side of \eqref{adgio} that can
be written as
$$
\sum_{\sigma\in \Gamma}\int_\Omega dx
\rho_\lambda(x,\sigma)g(x,\sigma)\left[F_{\sigma}(x)\cdot \nabla
f(x,\sigma)+\lambda\sum_{\sigma'\in
\Gamma}r(\sigma,\sigma'|x)\Big(f(x,\sigma')-f(x,\sigma)\Big)\right]\,.
$$
With a change of variable in the discrete sum and an integration by
parts in the mechanical  variable for which the boundary terms
disappear due to conditions \eqref{bondcond}, we obtain
\begin{multline*}
 \sum_{\sigma\in \Gamma}\int_\Omega dx f(x,\sigma)\left\{-\nabla\cdot
\Big(\rho_\lambda(x,\sigma)g(x,\sigma)F_{\sigma}(x)\Big)\right.
\\
\left.+\lambda\sum_{\sigma'\in
\Gamma}\Big(r(\sigma',\sigma|x)\rho_\lambda(x,\sigma')g(x,\sigma')-
r(\sigma,\sigma'|x)\rho_\lambda(x,\sigma)g(x,\sigma)\Big)\right\}\,.
\end{multline*}
Using now \eqref{inv} we get
\begin{eqnarray*}
& &\sum_{\sigma\in \Gamma}\int_\Omega dx
f(x,\sigma)\Big\{-\nabla\cdot
\Big(\rho_\lambda(x,\sigma)g(x,\sigma)F_{\sigma}(x)\Big)+
g(x,\sigma)\nabla\cdot\Big(\rho_\lambda(x,\sigma)F_{\sigma}(x)\Big)\Big.
\\
& &\left.+\rho_\lambda(x,\sigma)\lambda\sum_{\sigma'\in
\Gamma}\left(r(\sigma',\sigma|x)\frac{\rho_\lambda(x,\sigma')}
{\rho_\lambda(x,\sigma)}g(x,\sigma')-
r(\sigma',\sigma|x)\frac{\rho_\lambda(x,\sigma')}{\rho_\lambda(x,\sigma)}
g(x,\sigma)\right)\right\}\,,
\end{eqnarray*}
that finally becomes
$$
\sum_{\sigma\in \Gamma}\int_\Omega dx
\rho_\lambda(x,\sigma)f(x,\sigma)\Big[-F_{\sigma}(x)\cdot \nabla
g(x,\sigma)+\lambda\sum_{\sigma'\in
\Gamma}r(\sigma',\sigma|x)\frac{\rho_\lambda(x,\sigma')}{\rho_\lambda(x,\sigma)}
\Big(g(x,\sigma')-g(x,\sigma)\Big)\Big].
$$
This ends the proof of \eqref{adgiogen}.

\smallskip

Let us conclude this section with some comments.  We have showed
that the time--reversed process  is still a PDMP with reversed
vector fields   and  with probability  rates of   chemical
transitions given by \eqref{defr+}.
 Note that in general these rates
can be $\lambda$--dependent, thus implying that  the chemical part
of the adjoint generator can also be $\lambda$--dependent. Moreover,
the chemical part of the adjoint generator $L^+_c[x]$ is irreducible
for all $x \in \O$, due to the irreducibility of $L_c[x]$ and the
identity \eqref{defr+}. We call $\mu^+(\cdot |x)$ the
quasistationary measure associated to $L_c ^+[x]$ and we remark that
it can be $\lambda$ dependent. As already observed   $L_c^+[x]$ and
$L_c [x]$ are not adjoint operators w.r.t. quasistationary measure
$\mu(\cdot|x)$.
% namely $L_c^+[x]$ is not the chemica time--reversed version
%of the Markov chain with generator $L_c[x]$ and  initial
%distribution $\mu(\cdot|x)$.
 Another consequence of our results is the
following: if for any fixed $x\in \Omega$ the chemical part
$L_c[x]$ of the direct generator $L$ is reversible with respect to
the quasistationary measure $\mu(\cdot|x)$, then also the chemical
part of the adjoint process $L_c^+[x]$ is reversible with respect to
its quasistationary measure $\mu^+(\sigma|x)$ and moreover it holds
\begin{equation}
\mu^+(\sigma|x)=\frac{\rho_\lambda^2(x,\sigma)}{\mu(\sigma|x)
Z_\lambda(x)}\,, \label{quaqua+=inv}
\end{equation}
where $Z_\lambda(x)$ is the normalization constant.
 In order to justify \eqref{quaqua+=inv},
 one can argue as follows. The reversibility of $L_c[x]$   is equivalent to the the detailed balance condition
\begin{equation}
\mu(\sigma|x)r(\sigma,\sigma'|x)=\mu(\sigma'|x)r(\sigma',\sigma|x),
\ \ \ \ \forall \sigma,\sigma', \ \forall x\in \Omega.
\end{equation}
Due to  \eqref{defr+}, this relation is equivalent to
$$
\mu(\sigma|x)r^+(\sigma',\sigma|x)\frac{\rho_\lambda(x,\sigma')}{\rho_\lambda(x,\sigma)}
=\mu(\sigma'|x)r^+(\sigma,\sigma'|x)\frac{\rho_\lambda(x,\sigma)}{\rho_\lambda(x,\sigma')}\,,
$$
that can be written as
$$
\frac{\rho^2_\lambda(x,\sigma')}{\mu(\sigma'|x)}r^+(\sigma',\sigma|x)
=\frac{\rho^2_\lambda(x,\sigma)}{\mu(\sigma|x)}r^+(\sigma,\sigma'|x).
$$
This equation states that the rates $r^+$ at $x$ satisfy the
detailed balance condition with respect to a measure on $\Gamma$
proportional to $\frac{\rho^2_\lambda(x,\sigma)}{\mu(\sigma|x)}$.
The proportionality factor can depend on $x$ and on $\lambda$. This
is exactly the content of equation \eqref{quaqua+=inv}.

\bigskip

Finally, a comment about reversibility: due to our results, the
direct PDMP can be reversible only if all the vector fields
$F_{\sigma}$ are identically zero. In this case the mechanical state
remains constant and the model reduces to  the continuous time
Markov chain $\s(\cdot)$.

\section{One dimensional models with two chemical states}
\label{One} In this section we discuss in detail some  1D models
with two chemical states, for which it is possible to compute
explicitly the invariant measure. We consider separately the cases
of $\O$ interval and  1D torus. For simplicity we consider vector
fields without equilibrium points in $\O$, although the main ideas
presented here can be used in more general cases.

\subsection{Interval}
\label{1dbound} We take  $\Omega=(a,b)\subset \bbR$ and
$\Gamma=\left\{0,1\right\}$. In order to have some mixing and
confinement inside $\Omega$, we consider vector fields $F_0,F_1$
such that (i) $F_0(x)<0$ and $F_1 (x)>0$ for all $x \in (a,b)$, (ii)
$F_0(a)=F_1 (b)=0$. Because of the irreducibility of the Markov
chain associated to $L_c[x]$, the jump rates $r(0,1|x)$ and
$r(1,0|x)$ must be positive for all $x \in (a,b)$. The stationarity
equations \eqref{inv} are given by
\begin{equation}
\left\{
\begin{array}{ccc}
\lambda\bigl(\rho_\lambda(x,1)r(1,0|x)-\rho_\lambda(x,0)r(0,1|x)\bigr)&=
&\partial_x\bigl(\rho_\lambda(x,0)F_0(x)\bigr)\,,\\
\lambda\bigl(\rho_\lambda(x,0)r(0,1|x)-\rho_\lambda(x,1)r(1,0|x)\bigr)&=
&\partial_x\bigl(\rho_\lambda(x,1)F_1(x)\bigr)\,,\\
\end{array}
\right.\label{invoned}
\end{equation}
from which we obtain
$$
\partial_x\bigl(\rho_\lambda(x,0)F_0(x)+\rho_\lambda(x,1)F_1(x)\bigr)=0\,,
$$
and consequently
\begin{equation}
\rho_\lambda(x,0)F_0(x)+\rho_\lambda(x,1)F_1(x)=c. \label{rel}
\end{equation}
Due to the  boundary condition \eqref{bondcond}, we know that $c$
must be zero.  Then relation \eqref{rel} allows to solve equations
\eqref{invoned} by separation of variables, leading to
\begin{equation}
\rho_\lambda(x,i)=\frac{e^{-\lambda
\int_{x_*}^x\left(\frac{r(0,1|z)}{F_0(z)}+
\frac{r(1,0|z)}{F_1(z)}\right)dz}}{Z|F_i(x)|}, \ \ \ \ i=0,1\,,
\label{invoneexa}
\end{equation}
where $x_*$ is a generic element of $(a,b)$ and $Z$ is the
normalization constant, if it exists. Note that the boundary
condition \eqref{bondcond} is automatically satisfied, since by
construction the constant $c$ in \eqref{rel} is zero and by
assumption $F_0(a)=F_1 (b)=0$. As the reader can check,
\eqref{invoneexa} is the only solution of \eqref{inv} compatible
with the boundary condition \eqref{bondcond}. It corresponds to a
probability measure if and only if the normalization constant $Z$ is
well--defined. Suppose for example that  the jump rates vary in the
interval $(c_1,c_2)$ with $c_1,c_2>0$, while $F_1(x)$ and $F_0(x)$
are of order one near $a$ and $b$ respectively.
 Then it is simple to check that $Z$ is
well defined.

\medskip

Knowing the stationary measure $\rho_\l$, we can compute via
\eqref{defr+} the jump rates of the time--reversed process:
\begin{equation}
\left\{
\begin{array}{l}
r^+(0,1|x) = r(1,0|x)\frac{|F_0(x)|}{|F_1(x)|}\,,\\
r^+(1,0|x) =  r(0,1|x)\frac{|F_1(x)|}{|F_0(x)|}\,.\\
\end{array}
\right.\nonumber
\end{equation}
We point out that these rates do not depend on $\l$.  Moreover, we
observe that  in the reversed--time process the $\O$--confinement is
related  to \eqref{nonlomesso}, while in the direct process the
$\O$--confinement is related by \eqref{scabound}.

\subsection{Torus}\label{torus} We take
 $\Omega=\mathbb R/\mathbb Z$, $\Gamma=\left\{0,1\right\}$ and  periodic vector fields $F_0(x)$, $F_1(x)$.
 We assume $F_0(x)$ and  $F_1(x)$ to be nonzero. Moreover, in order to assure the irreducibility of
 the Markov chain associated to $L_c[x]$, we take positive jump rates $r(0,1|x)$ and $r (1,0|x)$. The equations for the
stationary measure are still \eqref{invoned} to which we have to add
the periodic boundary conditions
\begin{equation}
\rho_\lambda(0,i)=\rho_\lambda(1,i)\,, \ \ \ \ i=0,1\,. \label{pbc}
\end{equation}
Let us call
\begin{equation}
S(x):=\int_{0}^x\left(\frac{r(0,1|z)}{F_0(z)}+
\frac{r(1,0|z)}{F_1(z)}\right)\,dz\,,\qquad x \in
\bbR\,.\label{defS}
\end{equation}
It is easy to check that, for any constant $k$,
\begin{equation}
\left\{
\begin{array}{l}
\rho_\lambda(x,0):=\frac{k}{F_0(x)}\int_x^{x+1}\left[\frac{r(1,0|y)}{F_1(y)}e^{\lambda\left(S(y)-S(x)\right)}\right]\, dy\,,\\
\rho_\lambda(x,1):=\frac{k}{F_1(x)}\int_x^{x+1}\left[\frac{r(0,1|y)}{F_0(y)}e^{\lambda\left(S(y)-S(x)\right)}\right]\, dy\,,\\
\end{array}
\right.\label{solper}
\end{equation}
 are solutions of \eqref{invoned}, satisfying
the  boundary conditions \eqref{pbc}, since for any  $x,y\in \bbR$
it holds
\begin{equation}
S(y)-S(x)=S(y+1)-S(x+1)\,. \label{salt}
\end{equation}
Moreover, there exists a unique value of $k$ such that the
expressions in the r.h.s. of \eqref{solper} are positive functions,
satisfying the
 normalization condition
$$
\sum_{\sigma=0,1}\int_0^1\rho_\lambda(x,\sigma) dx=1\,.
$$
On the other hand, it is simple to check that the above solution is
the only probability measure satisfying \eqref{inv} and the periodic
boundary conditions. Hence, we have obtained  the unique invariant
measure of the model.

\medskip

Knowing the stationary measure $\rho_\l$, we can compute the jump
rates of the time--reversed process:
\begin{equation}
\left\{
\begin{array}{l}
r^+(0,1|x) = r(1,0|x)\frac{F_0(x)}{F_1(x)}
\frac{\int_x^{x+1}\left[\frac{r(0,1|y)}{F_0(y)}e^{\lambda
S(y)}\right]\,dy}
{\int_x^{x+1}\left[\frac{r(1,0|y)}{F_1(y)}e^{\lambda S(y)}\right]\, dy}\,,\\
r^+(1,0|x) =
r(0,1|x)\frac{F_1(x)}{F_0(x)}\frac{\int_x^{x+1}\left[\frac{r(1,0|y)}{F_1(y)}e^{\lambda
S(y)}\right]\, dy}
{\int_x^{x+1}\left[\frac{r(0,1|y)}{F_0(y)}e^{\lambda S(y)}\right]\,dy}\,.\\
\end{array}
\right.\label{genlam} \end{equation} Note that in general, the above
rates are $\l$--dependent. In order to exhibit cases of
$\l$--independent rates $r^+(\cdot,\cdot |x)$, let us consider the
{\it equilibrium condition}
\begin{equation}
\int_{0}^1\left(\frac{r(0,1|z)}{F_0(z)}+
\frac{r(1,0|z)}{F_1(z)}\right)dz=0\,.\label{eqtorus}
\end{equation}
We point out that this condition   can hold only if the vector
fields $F_0$ and $F_1$ have opposite sign and that it  is equivalent
to the fact that  $S$ is a periodic function of period one. We claim
that the rates $r^+(\cdot,\cdot |x)$ do not depend on $\lambda$ if
the above equilibrium condition is satisfied. In fact, in this case,
from the definition of $S$ and from its periodicity  we obtain that
\begin{equation}\label{vinello}
\int_x^{x+1}\left[\left(\frac{r(1,0|y)}{F_1(y)}+\frac{r(0,1|y)}{F_0(y)}\right)e^{\lambda
S(y)}\right]\, dy = \frac 1\lambda \left(e^{\lambda
S(x+1)}-e^{\lambda S(x)}\right)=0\,.
\end{equation}
%Due to \eqref{salt}   we know that the last expression is a constant
%$c(\l)$. Moreover,  $c(\l)=0$ if and only if the equilibrium
%condition holds.  In this last case,
From \eqref{genlam} and \eqref{vinello} we conclude that
\begin{equation}
\left\{
\begin{array}{l}
r^+(0,1|x) = -r(1,0|x)\frac{F_0(x)}{F_1(x)}\,,\\
r^+(1,0|x)=-r(0,1|x)\frac{F_1(x)}{F_0(x)}\,.\\
\end{array}
\right.\label{genlams}
\end{equation}
Remember that in this case the vector fields have opposite sign so
that the above rates are positive.

When condition \eqref{eqtorus} is violated then  rates
\eqref{genlam}  can be $\lambda$--dependent. In this case we can
study the asymptotic behavior of $\rho_\l$ as $\l \uparrow \infty$
using some classical results (see for example \cite{B}) that we
recall for the reader's convenience. Let $f$ and $\cS$ be smooth
real functions on the interval
 $[a,b]$.
 If $\cS(y)< \cS(a)$ for any $y\in (a,b]$ and   $\cS'(a)<0$, then   it
holds
\begin{equation}
\lim_{\lambda \to \infty}\frac{\int_a^bf(y)e^{\lambda
\cS(y)}\,dy}{e^{\lambda
\cS(a)}\lambda^{-1}}=-\frac{f(a)}{\cS'(a)}\,. \label{ris1}
\end{equation}
If there exists  $y^*\in (a,b)$ such that $\cS(y)< \cS(y^*)$ for any
$y\in (a,b)$ different from $y^*$ and moreover $\cS''(y^*)<0$, then
it holds
\begin{equation}
\lim_{\lambda \to \infty}\frac{\int_a^bf(y)e^{\lambda
\cS(y)}\,dy}{e^{\lambda \cS(y^*)}\lambda^{-\frac
12}}=f(y^*)\sqrt{-\frac{2\pi}{\cS''(y^*)}}\,. \label{ris2}
\end{equation}

Let us suppose that the function $S$ defined by \eqref{defS} is
regular and  that for any $x\geq 0 $ the maximum of the function $S$
in the closed interval $[x,x+1]$ is assumed in at most one point of
the open interval $(x,x+1)$. This fact is guaranteed if for example
$S(z_1)\neq S(z_2)$ for any pair of critical points $z_1$ and $z_2$
(i.e. such that $S'(z_1)=S'(z_2)=0$). If $F_0$ and $F_1$ have the
same sign, this is always true since the function $S$ is strictly
increasing or strictly decreasing. Moreover, note that in general
$x$ and $x+1$ cannot be both maximum points of the function $S$ on
$[x,x+1]$, since due to \eqref{salt} the identity $S(x)=S(x+1)$
would imply the periodicity of $S$.  In addition to the previous
assumptions, we require that $S''(y)<0$  if the maximum of $S$ on
the interval $[x,x+1]$ is reached at the internal point $y$. We
point out that by the methods discussed in \cite{B} more general
cases can be considered.

Given $x \geq 0$, we define
 $y(x)\in [x,x+1]$ as follows.  We require that $S(y(x))=\max_{y\in [x,x+1]}S(y)$. If the maximum
point in $[x,x+1]$ is unique, then $y(x)$ is univocally  determined
from the above condition. The only other possibility is that there
is a maximum point at the boundary and a maximum point in the
interior of the interval $[x,x+1]$. In this case, we define $y(x)$
as the unique maximum point inside $(x,x+1)$.

From the results \eqref{ris1} and \eqref{ris2} it is easy to derive
that
\begin{equation}
\left\{
\begin{array}{l}
r^+(0,1|x) = r(1,0|x)\frac{F_0(x)}{F_1(x)}
\frac{r(0,1|y(x))F_1(y(x))}
{r(1,0|y(x))F_0(y(x))}+o(1)\,,\\
r^+(1,0|x) = r(0,1|x)\frac{F_1(x)}{F_0(x)}
\frac{r(1,0|y(x))F_0(y(x))}
{r(0,1|y(x))F_1(y(x))}+o(1)\,,\\
\end{array}
\right.\label{genlamlmd}
\end{equation}
where by  $o(1)$ we indicate a  term which is infinitesimal as
$\lambda$ diverges. Note that the first terms in the r.h.s. of
\eqref{genlamlmd} are $\lambda$--independent.

Note that if    $y(x)\in \left\{x,x+1\right\}$, by the periodicity
of $F_\s (x)$ and $r (\s,\s'|x)$  formula \eqref{genlamlmd} reduces
to
\begin{equation}
\left\{
\begin{array}{l}
r^+(0,1|x) = r(0,1|x)+o(1)\,,\\
r^+(1,0|x) = r(1,0|x)+o(1)\,.\\
\end{array}
\right.\label{genlamlmd1}
\end{equation}
This is always true  if the  vector fields $F_0$ and $F_1$ have the
same sign, since in this case  the function $S$ is monotone. On the
other hand, if  $y(x)\in (x,x+1)$, then necessarily it holds
\begin{equation}\label{bacetto}
S'(y(x))=\frac{r(0,1|y(x))}{F_0(y(x))}+
\frac{r(1,0|y(x))}{F_1(y(x))}=0\,,
\end{equation}
and \eqref{genlamlmd} reduces to
\begin{equation}
\left\{
\begin{array}{l}
r^+(0,1|x) = -r(1,0|x)\frac{F_0(x)}{F_1(x)}
+o(1)\,,\\
r^+(1,0|x) = -r(0,1|x)\frac{F_1(x)}{F_0(x)}
+o(1)\,.\\
\end{array}
\right.\label{genlamlmd2}
\end{equation}

\medskip

We conclude this section by pointing out another consequence of the
equilibrium condition \eqref{eqtorus}, which we know to be
equivalent to the periodicity of the function $S$. Setting
$$ C(\l)= \int_0 ^1 \ \left[\frac{r(1,0|y)}{F_1(y)}e^{\l S(y)}\right]\,
dy\,,
$$
due to  \eqref{vinello}, equations \eqref{solper} read
\begin{equation}
\left\{
\begin{array}{l}
\rho_\lambda(x,0):=\frac{k C(\l)}{F_0(x)}e^{-\l S(x)} \,,\\
\rho_\lambda(x,1):=\frac{-k C(\l) }{F_1(x)}e^{-\l S(x)} \,.\\
\end{array}
\right.\label{solperino}
\end{equation}

\section{Exactly solvable models}
\label{exactly} \noindent In general it is difficult to obtain a
closed expression for the invariant measure of our PDMPs. In this
section we discuss a class of models for which this is possible and
the invariant measure has a special structure.

\subsection{General framework}
The exact solutions of the 1D models discussed  in the previous
section  suggest to look for invariant measures of the form
\begin{equation}
\rho_\lambda(x,\sigma)=c(\lambda)e^{-\lambda
S(x)}\rho(x,\sigma)\,,\label{guess}
\end{equation}
where $c(\lambda)$ is a normalization factor depending only on
$\lambda$, $S(x)$ is a function depending only on $x$ and
$\rho(x,\sigma)$ is a measure density  on $\Omega\times \Gamma$ non
depending on $\lambda$.

\medskip

 Let us discuss some consequences of \eqref{guess} (other aspects
 will be discussed
 later, when studying the fluctuations of our models). If the
  PDMP has   invariant measure of the form \eqref{guess}, then the
rates $r^+$ of the adjoint process do not depend on $\lambda$:
\begin{equation}
r^+(\sigma,\sigma'|x)=r(\sigma',\sigma|x)\frac{\rho(x,\sigma')}{\rho(x,\sigma)}\,.
\label{r+}
\end{equation}
In particular, the chemical part $L_c^+[x]$ of the adjoint generator
$L^+$ does not depend on $\lambda$. This implies that the
quasistationary measure $\mu^+(\cdot|x)$ does not depend on $\l$.
Moreover, if $L_c[x]$
 is reversible with
respect to the quasistationary measure $\mu(\cdot|x)$, then
\eqref{quaqua+=inv} becomes
\begin{equation}
\mu^+(\sigma|x)=\frac{\rho^2(x,\sigma)}{\mu(\sigma|x) Z(x)}\,.
\label{quaqua+=invnoL}
\end{equation}
%in which the parameter $\lambda$ disappear.

\medskip

Let us analyze now  when \eqref{guess} can be a solution of the
stationary equations \eqref{inv}. Examples will be discussed at the
end of this section. Inserting \eqref{guess} in \eqref{inv},
 we obtain that  for any $\sigma\in
\Gamma$ it must hold
\begin{eqnarray}
& & \lambda c(\lambda)e^{-\lambda S(x)}\sum_{\sigma'\in
\Gamma}\Big(\rho(x,\sigma')r(\sigma',\sigma|x)-\rho(x,\sigma)r(\sigma,\sigma'|x)\Big)=
\nonumber \\
&  & -\lambda c(\lambda)e^{-\lambda S(x)}\nabla S(x)\cdot
\Big(\rho(x,\sigma)F_{\sigma}(x)\Big)+c(\lambda)e^{-\lambda
S(x)}\nabla\cdot\Big(\rho(x,\sigma)F_{\sigma}(x)\Big)\,.
\end{eqnarray}
Dividing by nonzero terms we get
\begin{eqnarray}
& &\lambda \Big[\sum_{\sigma'\in
\Gamma}\Big(\rho(x,\sigma')r(\sigma',\sigma|x)-\rho(x,\sigma)r(\sigma,\sigma'|x)\Big)
+\nabla S(x)\cdot \Big(\rho(x,\sigma)F_{\sigma}(x)\Big)\Big]\\
& & \quad +\nabla\cdot\Big(\rho(x,\sigma)F_{\sigma}(x)\Big)=0\,.
\end{eqnarray}
In the above formula we have a first order polynomial in $\lambda$
and we have equality to zero for any value of $\lambda$ if and only
if the coefficients of the zero and first order terms   are equal to
zero separately:
\begin{equation}
\left\{
\begin{array}{l}
\nabla\cdot\Big(\rho(x,\sigma)F_{\sigma}(x)\Big) =  0\,,\\
\sum_{\sigma'\in
\Gamma}\Big(\rho(x,\sigma)r(\sigma,\sigma'|x)-\rho(x,\sigma')r(\sigma',\sigma|x)\Big)
= \nabla S(x)\cdot \Big(\rho(x,\sigma)F_{\sigma}(x)\Big)\,.
\end{array}\label{stazguess}
\right.
\end{equation}
The first   equation, one for any fixed $\sigma\in \Gamma$, can be
written as
\begin{equation}
\nabla \varphi (x,\sigma)\cdot F_\sigma(x)=-\nabla \cdot F_\sigma
(x)\,, \label{trasp}
\end{equation}
where we set $\varphi(x,\sigma)=\log \rho(x,\sigma)$ (here and below
we suppose that $\rho (x,\s)>0$ for all $(x,\s)\in \O\times\G$ due
to some mixing property of the system). Under the same assumption,
the second equation, one for any fixed $\sigma\in \Gamma$, can be
written in the equivalent form
\begin{equation}
\gamma(\sigma|x)-\gamma^+(\sigma|x)=\nabla S(x)\cdot
F_{\sigma}(x)\,. \label{eqeq}
\end{equation}

Equation \eqref{trasp} is a non--homogeneous transport equation
along the orbits of the vector fields $F_\sigma $. The important
fact is that, as $\s$ varies, these equations are uncoupled, hence
one can easily integrate them. We distinguish between closed and
open orbits.

\smallskip

{\it Closed orbits:} Let  $\gamma\subseteq \Omega$  be a closed
orbit of the vector field $F_\sigma$, let $x_0$ be any element of
$\gamma$ and let $x_\gamma(t)$ be the parametrization of $\gamma$
such that $x_\gamma(0)=x_0$ and
$\dot{x}_\gamma(t)=F_\sigma(x_\gamma(t))$ for all $t\in[0,T]$, where
$T$ is the period of the orbit (note that $x_\gamma(0)=x_\gamma
(T)=x_0$). Given $x'\in \gamma$,  let $t'$ be the only time in
$[0,T]$  such that $x'=x_\gamma(t')$. Then  \eqref{trasp} implies
that
\begin{multline} \label{integratrasp}
\varphi(x',\sigma)-\varphi(x_0,\sigma) =  \int_0^{t'}\nabla \varphi
(x_\gamma(s),\sigma)\cdot \dot{x}_\gamma(s)\, ds =\\
\int_0^{t'}\nabla \varphi (x_\gamma(s),\sigma)\cdot
F_\sigma(x_\gamma(s))\,ds =  -\int_0^{t'}\nabla \cdot
F_\sigma(x_\gamma(s))\, ds\,.
\end{multline}
This implies  that a function  $\varphi$ satisfying \eqref{trasp}
can be constructed along  $\gamma$ if and only if the univalued
condition
\begin{equation}
\int_0^{T}\nabla \cdot F_\sigma(x_\gamma(s))\, ds=0
\label{univalued}
\end{equation}
is satisfied. Moreover the values of $\varphi$ on $\gamma$ are
uniquely determined from \eqref{integratrasp} once the  initial
condition $\varphi(x_0,\sigma)$ has been arbitrarily fixed.

\smallskip

{\it Open orbits:} Let $\gamma\subseteq \Omega$ be an open orbit  of
the vector field $F_\sigma$ and consider $x_0\in \gamma$. Let
$x_\gamma(t)$ be a parametrization of $\gamma$ such that
$\dot{x}_\gamma(t)=F_\sigma(x_\gamma(t))$ and $x(0)=x_0$. Given
$x'\in \gamma$, let $t'$ be the unique time such that
$x_\gamma(t')=x'$. Then \eqref{integratrasp} continues to hold and
we can determine the value of $\varphi(x',\sigma)$ for any $x'\in
\gamma$ starting from the arbitrary initial condition
$\varphi(x_0,\sigma)$.
%We call $\partial \Omega_\sigma^+\subseteq
%\partial \Omega$ the set of boundary points from which orbits
%of the vector field $F_\sigma$ exit.

\medskip

Once  determined  $\rho(x,\sigma)$  by  integration of the first
group of equations in \eqref{stazguess},  we need to find a function
$S$ satisfying the second  group of equations in \eqref{stazguess}.
Let us call
\begin{equation}\label{dido}
D(x,\sigma):=\sum_{\sigma'\in
\Gamma}\Big(\rho(x,\sigma)r(\sigma,\sigma'|x)-\rho(x,\sigma')r(\sigma',\sigma|x)\Big)\,.
\end{equation}
Then, the second group of equations in \eqref{stazguess} reads
\begin{equation}\label{gianduia}
D(x,\s)= \nabla S (x) \cdot\bigl ( \rho (x,\s) F_\s (x) \bigr)\,,
\qquad \forall (x,\s) \in \O \times \G \,.
\end{equation}
This means that, for any $\sigma \in \Gamma$,  $D(x,\sigma)$ is the
directional derivative of $S$ at $x$ along the vector
$\rho(x,\sigma)F_\sigma(x)$. In fact, given a vector $v\in \mathbb
R^d$  the directional derivative of $S$  along $v$ at $x$ is given
by
\begin{equation}\label{stella}
\lim_{t\to 0}\frac{S(x+tv)-S(x)}{t}=\nabla S(x)\cdot v\,.
\end{equation}
Let us assume that there exists $\Gamma (x)\subseteq \Gamma$ with
the property that the vectors $ \rho(x,\sigma)F_\sigma(x)$,
$\sigma\in \Gamma(x)$, are linearly independent and satisfy
\begin{equation}
Span\left\{\rho(x,\sigma)F_\sigma(x)\right\}_{\sigma\in
\Gamma(x)}=Span\left\{\rho(x,\sigma)F_\sigma(x)\right\}_{\sigma\in
\Gamma}=\bbR^d \,, \label{span}
\end{equation}
where for generic vectors  $v_1, \dots ,v_k$  in $\bbR^d$ we write
$$
Span\left\{v_i\right\}_{i=1,\dots ,k}:=\Big\{v\in \mathbb R^d: \
v=\sum_{i=1}^k\lambda_iv_i \ , \lambda_i\in \mathbb R\Big\}\,.
$$
Note that \eqref{span} implies the identity $d=|\G(x)|$ and that,
since $\rho(x,\s)>0$, \eqref{span} is equivalent to the condition
\begin{equation}
Span\left\{F_\sigma(x)\right\}_{\sigma\in
\Gamma(x)}=Span\left\{F_\sigma(x)\right\}_{\sigma\in \Gamma}=\bbR^d
\,. \label{spanvero}
\end{equation}
%and we could have used the second group of equations in
%\eqref{stazguess} written in the equivalent form \eqref{eqeq}.

Due to \eqref{span},  for any $\sigma^*\not \in \Gamma(x)$ there
exist real numbers $c_\sigma(\sigma^*,x)$ such that
\begin{equation}\label{vito}
\rho(x,\sigma^*)F_{\sigma^*}(x)=\sum_{\sigma\in
\Gamma(x)}c_\sigma(\sigma^*,x)\rho(x,\sigma)F_\sigma(x)\,.
\end{equation}
Let us recall a trivial consequence of \eqref{stella}:
 given vectors $v,v_1,
\dots ,v_k$ such that $v=\sum_{i}c_iv_i$,  it holds
\begin{equation}
\lim_{t\to 0}\frac{S(x+tv)-S(x)}{t}=\sum_i c_i \nabla S(x)\cdot v_i=
\sum _i c_i\lim_{t\to 0}\frac{S(x+tv_i)-S(x)}{t} \,. \label{derdir}
\end{equation}
From \eqref{vito} and \eqref{derdir} we get that a necessary
condition for the existence of the function $S$ is given by the
family of identities
\begin{equation}
D(x,\sigma^*)=\sum_{\sigma\in
\Gamma(x)}c_\sigma(\sigma^*,x)D(x,\sigma) \ \ \ \ \ \forall
\sigma^*\not \in \Gamma(x). \label{condsufflin}
\end{equation}
%This condition follows from the fact that, given vectors $v,v_1,
%\dots ,v_k$ such that $v=\sum_{i}c_iv_i$,  it holds
%\begin{equation}
%\lim_{t\to 0}\frac{S(x+tv)-S(x)}{t}=\sum_i c_i \nabla S(x)\cdot
%v_i\,. \label{derdir}
%\end{equation}
We now derive a second necessary condition concerning $D$.
 Let
$\left\{A_{i,\sigma}(x)\right\}^{\sigma\in\Gamma(x)}_{i\in
\left\{1,\dots ,d\right\}}$ be the  $d\times d$ matrix, with  rows
labeled by the spatial dimensions and  columns labeled by the
chemical states in $\Gamma(x)$,  defined as follows. The column with
label $\sigma$ is given by the vector $\rho(x,\sigma)F_\sigma(x)$,
i.e. $A_{i,\sigma}(x)=\rho(x,\sigma)F_{\sigma}(x)\cdot e_i$, where
$(e_i, 1\leq i \leq d )$ denotes the canonical basis of $\bbR^d$.
This definition implies that
\begin{equation}\label{raul}
\rho(x,\sigma)F_\sigma(x)=\sum_{i=1}^dA_{i,\sigma}(x)e_i\,,\qquad
\forall \s \in \G(x)\,,
\end{equation}
as well as
\begin{equation}
e_i=\sum_{\sigma\in \Gamma(x)}A^{-1}_{
\sigma,i}(x)\rho(x,\sigma)F_\sigma(x)\,, \qquad \forall i: 1\leq i
\leq d\,. \label{cruc}
\end{equation}
The matrix $A^{-1}(x)$ is the inverse of the matrix $A(x)$ and has
columns labeled by the spatial dimensions, and rows labeled by the
chemical states in $\Gamma(x)$. Since
$\left\{D(x,\sigma)\right\}_{\sigma\in \Gamma(x)}$ must correspond
to directional derivatives of a function $S$ on $\Omega$, due to
\eqref{derdir} and \eqref{cruc} it must be
\begin{equation}
\partial_{x_i} S(x)=\sum_{\sigma\in
\Gamma(x)}A^{-1}_{\sigma,i}(x)D(x,\sigma)\,. \label{parS}
\end{equation}

Consider a closed curve $\gamma\subseteq \Omega$ with
parametrization $\left\{x_\gamma(t)\right\}_{t\in[0,T]}$. Since $\g$
is closed, it must be
\begin{equation}
0=\int_0^T\nabla S(x_\gamma(s))\cdot
\dot{x}_\gamma(s)\,ds\,.\label{condzero}
\end{equation}
Using \eqref{parS} the above identity becomes
\begin{equation}
0=\int_0^T\sum_{i=1}^d \Big(\sum_{\sigma\in \Gamma
(x_{\gamma}(s))}A^{-1}_{\sigma,i}(x_{\gamma}(s))D(x_{\gamma}(s),\sigma)\Big)\dot{x}_{\gamma,i}(s)\,
ds=\oint_\gamma \omega\,, \label{interocond}
\end{equation}
where $\omega$ is the differential form
\begin{equation}
\omega=\sum_{i=1}^d\Big(\sum_{\sigma\in
\Gamma(x)}A^{-1}_{\sigma,i}(x)D(x,\sigma)\Big)dx_i\,.
\label{diffform}
\end{equation}
The validity of condition \eqref{interocond} for any closed curve
$\gamma$ is equivalent to require that the differential form
\eqref{diffform} is exact.  When the domain $\Omega$ is simply
connected, exactness of $\o$  is equivalent to say that
\begin{equation}
\partial_{x_i}\Big(\sum_{\sigma\in
\Gamma(x)}A^{-1}_{\sigma,j}(x)D(x,\sigma)\Big)=\partial_{x_j}\Big(\sum_{\sigma\in
\Gamma(x)}A^{-1}_{\sigma,i}(x)D(x,\sigma)\Big), \ \ \ \ \forall
i\neq j. \label{derinc}
\end{equation}
%Note that summing over $\sigma\in \Gamma$ the left hand sides of all
%the equations of the second group in \eqref{stazguess} one gets
%zero. This means that

\medskip

We can now collect all our observations and reach a conclusion:
conditions \eqref{span} and \eqref{condsufflin} together with the
exactness of $\o$   allow
 to determine a solution $S$ of the second group of equations in
\eqref{stazguess}, up to additive terms, as follows.  Fix an
arbitrary point $x^*\in \Omega$ and a corresponding arbitrary value
$S(x^*)$. Given an arbitrary curve $\gamma$ that starts in $x^*$ and
ends in $x\in \Omega$ (recall that for simplicity we have taken $\O$
connected), the value $S(x)$ is given by
\begin{equation}
S(x)=\int_\gamma \omega+S(x^*)\,. \label{integro}
\end{equation}
Due to the exactness of $\omega$ this value does not depend on the
particular curve $\gamma$ chosen. Note that by construction
$dS=d\o$, and therefore \eqref{parS} is satisfied for any direction
$i$. Due to  \eqref{raul}, \eqref{cruc} and \eqref{derdir}, this
implies that $D(x,\s)$ is the directional derivative of $S$ along
the vector $\rho(x,\s)F_\s (x)$ at $x$, for any $x\in \O$ and for
any $\s \in \G(x)$. Due to condition \eqref{condsufflin} and to the
additivity property \eqref{derdir}, the same result extends also to
states $\s \in \G\setminus \G(x)$. Hence, the function $S$ satisfies
the second group of equations in \eqref{stazguess}.

\medskip

Above  we have   constructed solutions  of \eqref{inv} having  the
special form \eqref{guess}. To have that these solutions coincide
with the invariant measures of the PDMP, we have to impose the
boundary conditions \eqref{bondcond} if $\O$ is a bounded domain in
$\bbR^d$, or periodic boundary conditions in $x$  if $\O$ is the
$d$--dimensional torus.  This has to be done, when possible, using
the arbitrariness in the initial data in the above construction.

\medskip

We end our general discussion with a remark.  Since by definition
\eqref{dido} the sum   $\sum_{\sigma\in \Gamma}D(x,\sigma)$ must be
zero  and since by \eqref{stazguess} it holds  $D(x,\s)= \nabla S
(x) \cdot \bigl( \rho(x,\sigma)F_\sigma(x)\bigr)$,  it must be
\begin{equation}
\nabla S(x)\cdot \left(\sum_{\sigma\in
\Gamma}\rho(x,\sigma)F_\sigma(x)\right)=0\,. \label{ortentr}
\end{equation}
 In particular, the orbits of  the vector field $\sum_{\sigma\in
\Gamma}\rho(x,\sigma)F_\sigma(x)$ must lie inside the level curves
of $S$.

\subsection{Examples}
\label{Ex} We now exhibit solutions of \eqref{inv} in specific
examples  by means of the above construction. The reader can easily
check that our solutions satisfy the appropriate boundary conditions
so that they correspond to  the invariant measure of the PDMPs under
consideration.

\medskip

\subsubsection{Interval} We review  the PDMP of Subsection
\ref{1dbound} by means of the general method described above. We
keep the notation and assumptions already stated in Subsection
\ref{1dbound}. In particular, $\O=(a,b)$, $F_1$ has a unique orbit,
which is open and exits from $a$, while $F_0$ has a unique orbit,
which is open and  and exits from $b$. We fix an arbitrary point
$x_*\in (a,b)$.  Given $x(t)$ the solution of
\begin{equation}
\left\{
\begin{array}{l}
\dot{x}=F_1(x)\,,\\
x(0)=x_*\,,\\
\end{array}
\right.\label{ode1}
\end{equation}
then due to \eqref{trasp} for any $x\in (a,b)$ the function $\varphi
(x,1)= \log \rho (x,1)$ satisfies
\begin{equation}
\varphi(x,1)=\phi(x_*)-\int_0^t\nabla\cdot F_1(x(s))\,ds
\label{ins1}
\end{equation}
where the time $t$ is such that $x(t)=x$ and $\phi(x_*)$ is an
arbitrary constant. Differentiating \eqref{ode1} we get
$$
\frac{\ddot{x}(s)}{\dot{x}(s)}=\nabla \cdot F_1(x(s))
$$
that inserted in \eqref{ins1} gives
$$
\varphi(x,1)=\phi(x_*)-\int_0^t\frac{d}{ds}\left(\log
\dot{x}(s)\right)ds=\phi(x_*)+\log\frac{F_1(x_*)}{F_1(x)}\,.
$$
The above identity and similar arguments applied to the vector field
$F_0$ imply that
\begin{equation}\label{napoleone}
\begin{cases}
\rho(x,1)=e^{\phi(x_*)}\frac{F_1(x_*)}{F_1(x)}\,,\\
\rho(x,0)=e^{\psi(x_*)}\frac{F_0(x_*)}{F_0(x)}\,,
\end{cases}
\end{equation}
where also  $\psi(x_*)$ is  an arbitrary constant. It remains now to
determine the function $S$ and afterwards to fix the arbitrary
constants.  Taking  $\Gamma (x)=\left\{1\right\}$ for any $x\in
\Omega$, condition \eqref{spanvero} is  satisfied. Due to
\eqref{napoleone},  we can express the constant $c_1 (0,x)$ in
\eqref{vito} as \begin{equation}\label{antonio} c_1 (0,x) = \frac{
\rho (x,0) F_0(x) }{ \rho (x,1) F_1 (x) }=\frac{ e^{\psi (x_*)} F_0
(x_*) }{e^{\phi(x_*) } F_1 (x_*) }\,.
\end{equation}
Since in addition $D(x,0)=-D(x,1)$, condition \eqref{condsufflin} is
satisfied if and only if   $c_1(0,x) \equiv -1$. To this aim we take
$ e^{\psi(x_*)}=-1/F_0(x_*)$ and $ e^{\phi(x_*)}=1/F_1(x_*)$.
%up to an arbitrary additive constant that contribute to the
%normalization factor $c(\lambda)$.
By this choice, the differential form \eqref{diffform} is given by
\begin{equation}
\omega=\frac{D(x,1)}{\rho(x,1) F_1 (x) } dx=
\left(\frac{r(0,1|x)}{F_0(x)}+
\frac{r(1,0|x)}{F_1(x)}\right)dx\,.\label{diffform1b}
\end{equation}
The form $\o$ is trivially exact, being a 1D form  on a simply
connected domain. Then, by  formula \eqref{integro} and the previous
computations,
 we get \eqref{invoneexa} as a special
case of \eqref{guess}.

\subsubsection{1D torus} In the case of the 1D torus  discussed in Subsection \ref{torus}, we have $\Omega=\mathbb
R/\mathbb Z$ and both $F_0$ and $F_1$ have a closed orbit that
coincide with $\Omega$. In both cases the univalued condition
\eqref{univalued} is satisfied due to the fact that on the periodic
orbit $x(t)$ solution of $\dot{x}=F_i(x)$ with period $T$ we have
$$
\int_0^T\nabla \cdot F_i(x(s))\,ds=\int_0^T\frac{d}{ds}\left(\log
\dot{x}(s)\right)\,ds=0.
$$
We can repeat all the arguments and computations of the case
$\O=(a,b)$. The only exception is that now the form $\o$ given by
\eqref{diffform1b} is not automatically exact, since $\O$ is not
simply connected. The requirement that $ \oint_\Omega \omega=0
$
 is exactly the equilibrium condition \eqref{eqtorus}. The final result coincides with
 \eqref{solperino}.

\subsubsection{Triangular domain}. Let us now discuss a simple but
non trivial example in dimension $d=2$ of a PDMP with an invariant
measure of the form \eqref{guess}. Let  $\Omega\subseteq \mathbb
R^2$ be the  open triangle  with vertices $(0,0),(1,0),(0,1)$ and
let $(x,y)$ denote  a generic element of $\Omega$. The set of
chemical states is $\Gamma=\left\{1,2,3\right\}$. The vector fields
associated to the chemical states are obtained from the gradients of
quadratic potentials centered at the vertices of the triangle. More
precisely
\begin{equation}
\left\{
\begin{array}{l}
F_1(x,y)=-\frac 12\nabla \left(x^2+y^2\right)=(-x,-y)\,,\\
F_2(x,y)=-\frac 12\nabla \left((x-1)^2+y^2\right)=(1-x,-y)\,,\\
F_3(x,y)=-\frac 12\nabla \left(x^2+(y-1)^2\right)=(-x,1-y)\,.\\
\end{array}
\right.\nonumber
\end{equation}
All the orbits of the above  vector fields are open and condition
(\ref{scabound}) is satisfied. Moreover the orbits of the vector
fields $F_i$ exit from $\partial \Omega^+_i\subset \partial \Omega$,
where $\partial \Omega^+_1$ is the segment with extrema $(0,1)$ and
$(1,0)$; $\partial \Omega^+_2$ is the segment with extrema $(0,0)$
and $(0,1)$; $\partial \Omega^+_3$ is the segment with extrema
$(0,0)$ and $(1,0)$. Let us determine the function $\varphi (x,y,1)=
\log \rho(x,y,1)$ by means of the discussion following
\eqref{trasp}. To this aim, we observe that given $(x,y)\in \O$ the
path $\bigl(x(t),y(t)\bigr):=\bigl(e^{-t}, e^{-t}y/x\bigr)$
satisfies $$
\begin{cases} \bigl(\dot x(t), \dot y(t)\bigr)= F_1
\bigl(x(t),y(t)\bigr)\,, \qquad \forall t\geq t_0 \,,
\\
\bigl(x(t_0),y(t_0)\bigr)= (x/(x+y),y/(x+y))\in \partial \O_1^+\,, \\
\bigl(x(t_1),y(t_1)\bigr)= (x,y) \,,
\end{cases}
$$
where   $t_0:= \log \bigl( (x+y)/x\bigr)$ and  $t_1:= \log (1/x)$.
 In particular, the above path parameterized by $t \geq t_0$
  is an orbit of $F_1$ exiting from
$\partial \O_1^+$ and passing through the point $(x,y)$. Fixed an
arbitrary function $\phi_1:(0,1)\to \mathbb R$ we obtain using
\eqref{integratrasp}
$$
\varphi(x,y,1)-\phi_1\left(\frac{x}{x+y}\right)=2\int_{\log\frac{x+y}{x}}^{\log\frac{1}{
x}} dt=-2\log(x+y)\,,
$$
so that $
\rho(x,y,1)=\frac{e^{\phi_1\left(\frac{x}{x+y}\right)}}{(x+y)^2} $.
Note that this can be rewritten as $\rho ( x,y,1)=a_1 ( x/(x+y) )
/x^2$ for a suitable function $a_1:(0,1)\to \mathbb R$. In
conclusion, by similar arguments, we get that
\begin{equation}\label{balena}
\begin{cases}
\rho( x,y,1)= a_1\left( \frac{x}{x+y} \right) \frac{1}{x^2}\,,\\
\rho( x,y,2)= a_2\left( \frac{y}{1-x} \right) \frac{1}{y^2}\,,\\
\rho( x,y,3)= a_3\left( \frac{x}{1-y} \right) \frac{1}{x^2}\,,
\end{cases}
\end{equation}
for positive functions $a_1,a_2,a_3$, which can be chosen
arbitrarily. Note that the point $(0, y/(1-x))$ is the exit point in
$\partial\O_2^+$ of the $F_2$--orbit passing through the point
$(x,y)\in \O$, while $( x/(1-y) ,0)$ is the exit point in $\partial
\O _3^+$ of the $F_3$--orbit passing through the point $(x,y)\in
\O$.

 In order to determine the function $S$ of
\eqref{guess},
 for any $(x,y)\in \Omega$ we take $\Gamma(x,y)=\{2,3 \}$. Trivially
 condition \eqref{spanvero} is satisfied. Moreover, we can compute
 $c_2(1,x,y)$ and $c_3 (1,x,y )$ of \eqref{vito}:
\begin{equation}
\begin{cases}
c_2( 1,x,y )= \frac{ x}{x+y-1} \frac{\rho(x,y,1) }{\rho(x,y,2)}\,,
\\
c_3(1,x,y )= \frac{y}{x+y-1} \frac{ \rho(x,y,1) }{\rho(x,y,3)}\,.
\end{cases}
\end{equation}
At this point, the check of condition \eqref{condsufflin} depends
strongly from the form of the the rates $r(\cdot,\cdot| x,y)$.
Indeed, omitting   the dependence from the point $(x,y)$ (for the
sake of simplicity) condition \eqref{condsufflin} becomes
\begin{multline}\label{cavallo}
r(1,2)+r(1,3) -\frac{\rho(2)}{\rho(1)}
r(2,1)-\frac{\rho(3)}{\rho(1)} r(3,1) =\\ \frac{x}{x+y-1} \left[
r(2,1)+r(2,3) -\frac{\rho(1)}{\rho(2) } r(1,2)  -\frac{\rho(3) }{\rho (2) } r(3,2) \right]+\\
\frac{y}{x+y-1}  \left[ r(3,1)+r(3,2)-\frac{\rho(1)}{\rho(3)} r(1,3)
-\frac{\rho(2)}{\rho(3)} r(2,3) \right]\,.
\end{multline}
As the reader can easily check, the above identity is automatically
satisfied for all kind of jump rates if
\begin{equation}\label{libri} \frac{\rho(1)}{\rho(3) } =
\frac{1-x-y}{y} \,, \qquad \frac{\rho(2)}{\rho(3)}= \frac{x}{y}\,,
\qquad \frac{\rho(1)}{\rho(2)} = \frac{1-x-y}{x}\,.
\end{equation}

In order to satisfy the above identities it is enough to take
$a_1,a_2,a_3$ in \eqref{balena} as $a_i(u)= \frac{ u}{1-u}$. By this
choice, \eqref{balena} reads
\begin{equation}\label{balenabis}
\begin{cases}
\rho( x,y,1)= 1/(xy)\,,\\
\rho( x,y,2)= 1/[y(1-x-y)] \,,\\
\rho( x,y,3)= 1/[x(1-x-y)]\,.
\end{cases}
\end{equation}
%Note that by this choice, the boundary condition \eqref{bondcond} is
%satisfied.
It remains now to compute the form $\o$ given by \eqref{diffform},
check when it is exact and afterwards check the boundary condition
\eqref{bondcond}. First we observe that
$$
A(x,y)= \frac{1}{1-x-y} \left(\begin{array}{cc}
\frac{1-x}{y} & -1\\
-1 & \frac{1-y}{x}\\
\end{array}
\right)\,, \qquad A^{-1}(x,y)=  \left(\begin{array}{cc}
y(1-y) & xy\\
xy & x(1-x)\\
\end{array}
\right)\,.
$$
Therefore
$$ \o = \left[ y(1-y) D(x,y,2) +xy D(x,y,3)\right]dx + \left[ xy D(x,y,2)+x(1-x) D(x,y,3)\right]dy\,.$$
%By omitting somewhere the dependence on $(x,y)$
 We have
$$ \o= B(x,y) dx +C(x,y) dy$$
where (omitting the dependence on $(x,y)$ for simplicity)
\begin{align*} & B(x,y)= \frac{1-y}{1-x-y} r(2,1)-\frac{1-y}{x}
r(1,2)+ \frac{y}{1-x-y} r(3,1)-r(1,3)+ r(2,3)-
\frac{y}{x }r(3,2)\,,\\
& C(x,y)= \frac{1-x}{1-x-y} r(3,1)-\frac{1-x}{y} r(1,3)+
\frac{x}{1-x-y} r(2,1)-r(1,2)+ r(3,2)- \frac{x}{y }r(2,3)\,.
\end{align*}
Note that $C(x,y)$ can be obtained from $B(x,y)$ by exchanging $x$
with $ y$ and $2$ with $3$.

If, motivated by the geometric symmetries of $\O$, we assume that
$$
r(1,2 |x,y)= r(1,3|y,x)\,, \qquad r(2,1|x,y)=r(3,1|y,x)\,, \qquad
r(2,3|x,y)=r(3,2|y,x)$$ then
$$ \o = B(x,y) dx + B(y,x) dy\,.$$
In this case, since $\O$ is simply connected, $\o$ is exact if and
only if $\partial_y B(a,b)=\partial _y B( b,a)$ (cf.
\eqref{derinc}), i.e the function $\partial_yB$ is symmetric.

Let us discuss an example, where the above condition is satisfied.
We take $r(\s,\s'|x,y)=1$ for all $\s \not = \s'$. Then one easily
compute the above $B(x,y)$ and $C(x,y)$,  getting
\begin{equation}
\omega=\frac{2x+y-1}{x(1-x-y)}dx+\frac{2y+x-1}{y(1-x-y)}dy\,,
\end{equation}
which is exact since the  domain $\Omega$ is simply connected and
condition \eqref{derinc} is satisfied. Integrating the form $\o$ as
in  \eqref{integro},  we obtain up to an arbitrary constant
\begin{equation}
S(x,y)=\int_\gamma \omega=-\log x -\log y -\log(1-x-y)\,.
\label{entr}
\end{equation}
By collecting our results \eqref{balenabis} and \eqref{entr}, we
obtain that the invariant measure is of the form \eqref{guess} and
that
\begin{multline*}
\rho_\lambda=\Big(\rho_\lambda(x,y,1),\rho_\lambda(x,y,2),
\rho_\lambda(x,y,3)\Big)=\nonumber \\
 c(\lambda)
\left(x^{\lambda-1}y^{\lambda-1}(1-x-y)^{\lambda},x^{\lambda}
y^{\lambda-1}(1-x-y)^{\lambda-1},x^{\lambda-1}y^{\lambda}(1-x-y)^{\lambda-1}\right)\,.
\end{multline*}
Above, $c(\l)$ is the normalization constant, which is well--defined
as the reader can easily check. Finally, we observe that the above
invariant measure satisfies the boundary condition \eqref{bondcond}.

Note that, due to \eqref{balena}, \eqref{bondcond} can be satisfied
only if $S$ diverges to $-\infty$ when approaching the boundary of
the triangle $\O$. This is a strong restriction. For example, if we
tale $r(3,2|x,y)=r(2,3|x,y)=0 $ , $r(2,1|x,y)=r(3,1|x,y)=1-x-y$,
$r(1,2|x,y)=r(1,3|y,x)=x$, we obtain that $\o= dS$ where $S$ is a
constant function, thus leading to a solution of the stationary
equations \eqref{inv}, but not satisfying the boundary condition
\eqref{bondcond}.

\subsubsection{2D torus} We take  $\Omega=\mathbb
R^2/\mathbb Z^2$ and $\Gamma=\left\{0,1\right\}$. We call $(x,y)$ a
generic element of $\Omega$ and choose  vector fields
\begin{equation}
\left\{
\begin{array}{l}
F_0(x,y)=(f(x,y),0)\,, \\
F_1(x,y)=(0,g(x,y))\,, \\
\end{array}
\right.\nonumber
\end{equation}
where $f$ and $g$ are regular  functions which never vanish on
$\Omega$. The chemical part of the generator is determined by the
transition rates $r(i,i-1|x,y)$ $i=0,1$. The first  group of
equations in \eqref{stazguess} can be  easily solved:
\begin{equation}
\left\{
\begin{array}{l}
\rho(x,y,0)=\frac{\phi(y)}{f(x,y)}\,, \\
\rho(x,y,1)=\frac{\tilde{\phi}(x)}{g(x,y)}\,; \\
\end{array}
\right.\nonumber
\end{equation}
where $\phi$ and $\tilde{\phi}$ are arbitrary functions. Moreover we
have that
\begin{equation}
A^{-1}(x,y)=\left(
\begin{array}{cc}
\frac{1}{\phi(y)} & 0 \\
0 & \frac{1}{\tilde{\phi}(x)} \\
\end{array}
\right)\,,\nonumber
\end{equation}
%where the lines are labeled respectively by the chemical states $0$
%and $1$ and the columns are labeled respectively by the spatial
%dimensions $x$ and $y$. We get the following differential form
hence $\o$ can be written as
\begin{equation}
\omega=\left(\frac{r(0,1|x,y)}{f(x,y)}-\frac{\tilde{\phi}(x)r(1,0|x,y)}{\phi(y)g(x,y)}\right)dx+
\left(\frac{r(1,0|x,y)}{g(x,y)}-\frac{\phi(y)r(0,1|x,y)}{\tilde{\phi}(x)f(x,y)}\right)dy\,.
\label{difform2d}
\end{equation}
Fix  arbitrary points $x^*,y^*\in [0,1]$ and consider the associated
fundamental cycles on $\Omega$
\begin{equation}
\left\{
\begin{array}{lc}
\gamma^1(t)=(t,y^*) & t\in [0,1]\,,\\
\gamma^2(t)=(x^*,t) & t\in [0,1]\,.\\
\end{array}
\right.\nonumber
\end{equation}
The exactness of \eqref{difform2d} is equivalent to impose
conditions \eqref{derinc} with the additional conditions
\begin{equation}
\oint_{\gamma^1} \omega=\oint_{\gamma^2} \omega=0\,.
\end{equation}
If we call $H(x,y)=\phi(y)r(0,1|x,y)/f(x,y)$ and $G(x,y)=
\tilde{\phi}(x)r(1,0|x,y)/g(x,y)$, then we can write $\o$ as
$$ \o = \frac{  H(x,y)-G(x,y)}{\phi(y) } dx + \frac{ G(x,y)-H(x,y) }{\tilde \phi (x)}dy
\,,
$$
hence
 the above exactness conditions
become
\begin{equation}
\left\{
\begin{array}{l}
\int_0^1H(x^*,y)dy=\int_0^1G(x^*,y) dy\,,\\
\int_0^1H(x,y^*)dx=\int_0^1G(x,y^*) dx,\\
\partial_x\left(\frac{G(x,y)-H(x,y)}{\tilde{\phi}(x)}\right)=
\partial_y \left(\frac{H(x,y)-G(x,y)}{\phi(y)}\right)\,.\\
\end{array}
\right.\nonumber
\end{equation}
Examples of rates $r(i,i-1|x,y)$ satisfying these conditions can be
easily constructed.

\subsubsection{Square domain} We consider  the open square
$\Omega \subseteq \mathbb R^2$  with vertices $(0,0)$, $(0,1)$,
$(1,0)$ and $(1,1)$. The chemical states are
$\Gamma=\left\{0,1,2,3\right\}$ with associated  vector fields
$$
\begin{array}{ll}
F_0(x,y)=(-x,-y)\,, & F_1(x,y)=(1-x,-y)\,,\\
F_2(x,y)=(-x,1-y)\,, & F_3(x,y)=\alpha (1-x, 1-y)\,,\\
\end{array}
$$
where $\alpha$ is a positive parameter and $(x,y)$ is a generic
element of $\Omega$. We choose the jump rates as
$$
\begin{array}{ll}
r(0,1|x,y)=r(0,2|y,x)=q(x,y)\,, & r(1,0|x,y)=r(2,0|y,x)=r(x,y)\,,\\
r(1,3|x,y)=r(2,3|y,x)=Q(x,y)\,, & r(3,1|x,y)=r(3,2|y,x)=R(x,y)\,,\\
\end{array}
$$
where $q$, $r$, $Q$ and $R$ are arbitrary positive functions and
moreover $r(1,2|x,y)=r(2,1|x,y)=0$.

Proceeding as in the previous examples we obtain a solution of the
form \eqref{guess} if we require that the rates satisfy the
following relations:  there exists a function $G(x,y)$ such that
$$
\left\{
\begin{array}{l}
q(x,y)-Q(x,y)=G(x,y)x\,,\\
r(x,y)-\frac{R(x,y)}{\alpha}=G(x,y)(1-x)\,.\\
\end{array}
\right.
$$
and there exists a symmetric function $s(x,y)$ and a function $\phi$
such that
$$
xR(x,y)-\alpha(1-x)Q(x,y)=\alpha x(1-x)\left(\int_z^y du \,
s(x,u)+\phi(x)\right)\,,
$$
where $z\in(0,1)$.  Under the above conditions we have a solution of
the form \eqref{guess} with
$$
S(x,y)=\int_{z}^x dw \int_z^y du\, s(w,u)+\int_z^x dw\,
\phi(w)+\int_z^y du \, \phi(u)\,,
$$
and
$$
\begin{array}{ll}
\rho(x,y,0)=\frac{1}{xy}\,, & \rho(x,y,1)=\frac{1}{y(1-x)}\,,\\
\rho(x,y,2)=\frac{1}{x(1-y)}\,, & \rho(x,y,3)=\frac{1}{\alpha (1-x)(1-y)}\,.\\
\end{array}
$$
Boundary conditions \eqref{bondcond} are not necessarily satisfied.

\section{Averaging  and large deviation principles in the high frequency limit}
\label{HFLLD} In this section we study the asymptotic  behavior of
our PDMPs and the corresponding time--reversed versions as  the
parameter $\lambda$ diverges to infinity. By this limit,  the
frequency of chemical jumps diverges and the timescale of chemical
jumps becomes infinitesimal w.r.t. the relaxation time of the
mechanical state.
 Below, we recall some rigorous results derived
in \cite{FGRmat}, where the interested reader can find  a more
detailed discussion.

In order to describe the  asymptotic behavior of the system  and
analyze deviations from it, we need to specify carefully both the
limit procedure and the involved spaces. Given a time interval
$[0,T]$, a trajectory $\bigl\{ x(t)\bigr\}_{t\in [0,T]}$ of the
mechanical variable is a continuous function $x:[0,T]\rightarrow
\Omega$. It is then natural to consider the  mechanical trajectories
as elements of the  path space $C([0,T], \Omega)$ endowed with the
topology induced by the $\sup$ norm. A chemical trajectory
$\left\{\sigma(t)\right\}_{t\in [0,T]}$ is an element of the
Skorokhod space $D([0,T],\Gamma)$ of right continuous functions
having left limit and taking values in $\Gamma=\{\s_1, \dots,
\s_{|\G|}\}$. To a chemical trajectory  we associate the following
time--dependent $d$--dimensional vector
$$
\bigl\{\sigma(t)\bigr\}_{t\in [0,T]}\to
\bigl\{\chi(t)\bigr\}_{t\in[0,T]}=\left\{\,
\bigl(\,\chi_{\sigma_1}(t), \dots
,\chi_{\sigma_{|\Gamma|}}(t)\bigr)\,\right\}_{t\in[0,T]}\,,
$$
%The set of chemical states is $\Gamma=\left\{\sigma_1,\dots
%,\sigma_{|\Gamma|}\right\}$.
%where  $\chi_{\sigma}$ denotes the characteristic function
where
\begin{equation*}
\chi_{\sigma}(t)=
\begin{cases}
1 & \text{  if  } \sigma (t)=\sigma\,,\\
0 & \text{ if } \sigma (t) \neq \sigma\,.\\
\end{cases}
\end{equation*}
We denote by
 $\mathcal M([0,T])$ the space  of nonnegative
finite measures on the interval $[0,T]$, endowed of the weak
convergence topology. Namely, $\mu_n \rightarrow \mu$ in $\mathcal
M([0,T])$ if and only if $\int _0^T f(t) \mu_n(dt)\rightarrow
\int_0^T f(t) \mu(dt) $ for all continuous functions $f$ on $[0,T]$.
Then we isolate the subspace $\cM_0([0,T])\subset \cM ([0,T]) $
given by the measures that are absolutely continuous w.r.t. the
Lebesgue measure. We can interpret
$\left\{\chi(t)\right\}_{t\in[0,T]}$ as an element of the cartesian
product $\mathcal M_0([0,T])^{\Gamma}$ by identifying $\{
\chi_\s(t)\}_{t \in [0,T]}$ with the measure $\chi_\s(t) dt$. If our
PDMP starts in the state $(x_0,\s_0)$, we can think its evolution
$\{ x(t), \chi (t) \}_{t \in [0,T]}$  as an element of following
subset $\cY_{x_0}$  of $C([0,T], \Omega)\times\mathcal
M_0([0,T])^{\Gamma}$:
\begin{multline}
\mathcal Y_{x_0}=\Big\{\left\{x(t),\chi(t)\right\}_{t\in [0,T]}\in
C([0,T], \Omega)\times\mathcal M_0([0,T])^{\Gamma} :\\
\sum_{\sigma\in \Gamma}\chi_{\sigma}(t)=1\ a.e.,\
x(t)=x_0+\int_0^t\sum_{\sigma\in
\Gamma}\chi_\sigma(s)F_{\sigma}(x(s))\,ds\Big\} \,.\label{spazio}
\end{multline}
Above, as in the rest of the paper, we write  $\{ x(t), \chi (t)
\}_{t \in [0,T]}$ instead of $\{ (x(t), \chi (t)) \}_{t \in [0,T]}$
in order to simplify the notation. Moreover,
  in the above formula
and hereafter we identify measures in  $\cM_0 ([0,T])$ with their
corresponding densities.
 It can be proved (cf. \cite{FGRmat}) that $\cY_{x_0}$ is a compact subspace of $C[0,T]\times \cM[0,T]^\G$,
and its topology can be derived from the metric $d$ defined as
\begin{multline}\label{sara}
d\left( \,\{x(t), \chi(t)\}_{t \in [0,T]}  \,,\, \{\bar x(t), \bar
\chi(t)\}_{t \in [0,T]} \,\right) =\\  \sup_{t \in [0,T]} |x(t)-\bar
x(t) | +\sum _{\s \in \G}  \sup _{0\leq t \leq T} \Bigl|\int _0 ^t
\bigl[ \chi_\s(s) -\bar \chi_\s (s) \bigr]ds \Bigr|\,.
\end{multline}
Moreover, in \cite{FGRmat} we prove the following law of large
numbers.
 Given $(x,\s) \in \O \times \G$, we define the mean vector  field
 $\bar F (x)$ as the average   with respect to
 the
 quasistationary measure $\mu(\cdot|x)$
  of the fields $F_\s(x)$:
\begin{equation}\label{campomedio}
  \bar F(x) = \sum _{\s \in \G} \mu(\s|x) F_\s(x) \,.
\end{equation}
Given the  initial state $(x_0,\sigma_0)$, we   call
$\left\{x^*(t),\chi^*(t)\right\}_{t\in [0,T]}$ the unique element of
$\mathcal Y_{x_0}$ such that
\begin{equation}
\left\{
\begin{array}{l}
\dot{x}^*(t)=\bar F (x^*(t))\,,\\
x^*(0)=x_0\,,\\
\chi^*_\sigma(t)=\mu(\sigma|x^*(t))\,.
\end{array}
\right.\label{idro}
\end{equation}
Then, the   following law of large numbers holds:
\begin{equation}
\lim_{\lambda \to \infty}\mathbb P^\lambda_{
x_0,\sigma_0}\Big[d\bigl(\left\{x(t),\chi(t)\right\}_{t\in
[0,T]},\left\{x^*(t),\chi^*(t)\right\}_{t\in
[0,T]}\bigr)>\delta\Big]=0\,, \ \ \ \forall \delta >0\,,
\end{equation}
where  the law   $\mathbb P^\lambda_{x_0,\sigma_0}$ of the PDMP
starting at $(x_0,\s_0)$ and having parameter $\l$ is thought of as
a probability distribution on  $\mathcal Y_{x_0}$. Above
$\left\{x(t),\chi(t)\right\}_{t\in [0,T]}$ denotes a typical element
of $\cY_{x_0}$. Finally, we point out that the limit element
$\left\{x^*(t),\chi^*(t)\right\}_{t\in [0,T]}$ is independent from
the initial chemical state $\sigma_0$.

The above law of large numbers  is a typical  example of  {\it
Averaging Principle}. Indeed, we are dealing with a stochastic
dynamical systems with  fully--coupled fast and slow variables. In
the high frequency limit,  the  fast variables $\chi$ average
according to the local quasistationary measure as the slow variables
$x$ would be frozen, while   the slow variables $x$ feel the
averaged vector field $\bar F$.

\smallskip

We briefly illustrate a {\it Large Deviation Principle}, where the
probability of deviations from   the above law of large numbers is
computed on exponential scale.  For  precise statements and rigorous
proofs we refer to \cite{FGRmat}, while we keep here  the exposition
 at a more heuristic level. We are interested in the exponential
 probability
rate   of rare events, namely we look for a functional $J_{[0,T]}$
on $\mathcal Y_{x_0}$ such that, for any fixed path  $\bigl\{ \hat x
(t), \hat \chi (t) \bigr\}\in \cY_{x_0}$, it holds
\begin{equation}
\mathbb
P^\lambda_{x_0,\sigma_0}\left(\left\{x(t),\chi(t)\right\}_{t\in
[0,T]}\approx \left\{\hat{x}(t),\hat{\chi}(t)\right\}_{t\in
[0,T]}\right)\sim e^{-\lambda J_{[0,T]}\bigl(
\bigl\{\hat{x}(t),\hat{\chi}(t)\bigr\}_{t\in[0,T]}\bigr)}\,.
\end{equation}
In the above formula, $\bigl\{x(t),\chi(t)\bigr\}_{t\in[0,T]}$ is a
generic element of $\mathcal Y_{x_0}$, the symbol $\approx$ means
closeness in the metric of $\mathcal Y_{x_0} $ and finally $\sim$
means asymptotic logarithmic equivalence in the limit of diverging
$\l$. The functional $J_{[0,T]}$ is called the rate functional.

For  our PDMPs  such a functional exists and has a variational
representation. In order to describe it, we fix some notation. We
denote by $W$ the set of pairs
$$ W:=\{ (\s,\s' ) \in \G \times \G\,:\, \s \not = \s'\}\,.$$
%Given a nonnegative measure $\p$ on $\G$ and nonnegative numbers
%$r(\s,\s')$, with $(\s,\s')\in W$, we define the function $j_*(\p,
%r)$ as
Given a point $x \in \O$ and a   vector $\chi \in [0,1]^\G$, we
define
\begin{equation}\label{pranzo}
j(x, \chi ):=\sup _{z\in (0,\infty)^\G} \sum _{(\s,\s')\in W }
\chi_\s  r(\s,\s'|x)\left [ 1-\frac{z_{\s'}}{z_\s } \right] \,.
\end{equation}
Then the rate functional $J_{[0,T]}:\cY_{x_0} \rightarrow
[0,\infty)$ is given by
\begin{equation}\label{ldp}
J_{[0,T]}\bigl(\bigl\{x(t),\chi(t)\bigr\}_{t\in[0,T]}\bigr):= \int
_0^T j \bigl ( x (t) ,\chi (t)  \bigr) dt\,.
\end{equation}
Note that the above functional does not depend on $\s_0$, but
depends on $x_0$ since its  domain is given by  $\cY_{x_0}$.

 If,
given $x \in \O$, the chemical part $L_c [x] $ of the generator is
reversible w.r.t. the quasistationary measure $\mu(\sigma|x)$, then
one  can solve the variational problem \eqref{pranzo} (see
\cite{FGRmat} for dettails) getting:
\begin{equation}
 j(x,\chi) =  \sum_\sigma
\gamma(\sigma|x)\chi_\sigma-  \sum _{(\s,\s')\in W}
\sqrt{\frac{\mu(\sigma|x)}{\mu(\sigma'|x)}}r(\sigma,\sigma'|x)
\sqrt{\chi_\sigma}\sqrt{\chi_{\sigma'}}\,. \label{density}
\end{equation}
We know that the above condition is always satisfied if $|\G|=2$.
Writing $\G=\{0,1\}$ one easily computes $j(\s,\chi)$ as
\begin{equation}\label{due}
j(x,\chi)= \left( \sqrt{ \chi_0 r(0,1|x) } -\sqrt{ \chi _1 r(1,0|x)}
\right)^2 \,.
\end{equation}

\subsection{LDP for the mechanical state}
 It is natural to analyze the statistical behavior of the
mechanical variables  alone, since often the  chemical variables
remain hidden to  direct observations.  To this aim, by means of the
contraction principle \cite{DZ},  one can derive the  LDP rate
functional $J^m_{[0,T]}: C([0,T],\O)\rightarrow [0,\infty]$ for the
 mechanical variables   from the joint (chemical and mechanical)
rate functional $J_{[0,T]}$ defined above. In particular,  given an
element $\{\hat x(t)\}_{t\in [0,T]}\in C([0,T],\Omega)$, for each
initial state $(x_0,\s_0)$ it holds
$$
\bbP ^\l _{x_0,\s_0} \left( \{ x(t)  \}_{t \in [0,T]} \approx \{
\hat x(t) \}_{t \in [0,T]} \right) \sim e^{- \l J^m_{[0,T]} \left(
\{ \hat x(t) \}_{t\in [0,T]}
\right)}\,,
$$
where
\begin{equation}
J_{[0,T]}^m\bigl(\{x(t) \}_{t\in [0,T]}\bigr)=\inf_{\left\{
\{\chi(t) \}_{t\in[0,T]}\,:\, \{x(t),\chi(t) \}_{t\in[0,T]}\in
\mathcal Y_{x_0}
\right\}}J_{[0,T]}\left(\{x(t),\chi(t)\}_{t\in[0,T]}\right)\,.
\end{equation}
Above, we have used the convention that the infimum over the empty
set is defined as  $+\infty$.
 From  expression \eqref{ldp} we
obtain that the functional $J^m_{[0,T]}:C([0,T], \O)\rightarrow
[0,\infty]$ equals
\begin{equation}
J_{[0,T]}^m\left(\{x(t)\}_{t\in [0,T]}\right)=
\begin{cases}
\int_0^Tj_m (x(t),\dot{x}(t))\,dt & \text{ if } x(\cdot) \in
\cY_{x_0}^m \,, \\
+\infty  & \text{ otherwise}\,, \end{cases} \label{LDPm}
\end{equation}
where
$$ \cY^m_{x_0} := \Big\{ \{x(t)\}_{t \in [0,T]} \,:\, \exists
\{\chi (t)\}_{t\in [0,T]}  \text{ s.t. }  \{x(t), \chi (t)\}_{t\in
[0,T]}  \in \cY _{x_0}\Big\}\,.
$$
and  the density $j_m(x,\dot{x})$ is given by
\begin{equation}
j_m (x,\dot{x})=\inf_{\left\{\chi:\ \dot{x}=\sum_\sigma \chi_\sigma
F_{\sigma}(x)\right\}}j(x,\chi)\,.\label{densitym}
\end{equation}
In the above formula, $\chi$ varies among vectors in $[0,1]^\G$ such
that $\sum_\s \chi_\s =1$.
%Here also, as in section \ref{exactly},
%for simplicity we restrict our discussion to models satisfying
%condition
%\begin{equation}
%Span\left\{F_\sigma(x)\right\}_{\sigma\in \Gamma}=\mathbb R^d\,, \ \
%\ \ \ \forall x\in \Omega\,.\label{spam}
%\end{equation}

In general an explicit computation of $j_m$ depends on the specific
model we are dealing with. In formula \eqref{densitym} we are
minimizing over all possible convex decompositions of the vector
$\dot x$ with respect to the collection of vectors
$\left\{F_\sigma(x)\right\}_{\sigma\in \Gamma}$. A special case is
when for any $x\in \Omega$ the collection of vectors
$\left\{F_\sigma(x)\right\}_{\sigma\in \Gamma}$ are the vertices of
a   simplex, i.e. the vectors  $\{ F_{\s_j}-F_{\s_1}\,:\, 2 \leq j
\leq |\G|\}$ are independent (writing  $\G=\{ \s_j :1\leq j \leq
|\G|\}$).
% (and then necessarily from
%\eqref{spam} we have $|\Gamma|= d$ or $|\Gamma|= d+1$).
 In this case
if the vector $\dot{x}$ belong to $\mathcal C
\left(\left\{F_\sigma(x)\right\}_{\sigma\in \Gamma}\right)$, where
the symbol $\mathcal C(\cdot)$ denotes the {\it convex hull}, then
there exists a unique probability measure on $\Gamma$,
$\chi^F(\dot{x})$, such that $\dot{x}=\sum_\sigma
\chi^F_\sigma(\dot{x})F_{\sigma}(x)$. The $\chi^F(\dot{x})$ are
called the {\it barycentric coordinates} of $\dot{x}$ with respect
to the collection of vectors $\left\{F_\sigma(x)\right\}_{\sigma\in
\Gamma}$. The upper index $F$ indicates the dependence on the vector
fields, the dependence on $x$ is understood. When $\dot x\not \in
\mathcal C \left(\left\{F_\sigma(x)\right\}_{\sigma\in
\Gamma}\right)$ then the infimum in \eqref{densitym} is over an
empty set and we obtain
\begin{equation}
j_m(x,\dot{x})=
\begin{cases}
j(x,\chi^F(\dot{x})) & \text{ if } \dot{x}\in \mathcal
C\left(\left\{F_\sigma(x)\right\}_{\sigma\in \Gamma}\right)\,,\\
+\infty & \text{ otherwise}\,.
\end{cases}
\label{densitymsimplex}
\end{equation}
We will compute the rate density $j_m (x,\dot{x})$ in specific
examples in the next section.

\subsection{LDP for the time--reversed process}
Since the adjoint  (time--reversed) process of our  PDMP is again a
PDMP with reversed vector fields,
 the space on which it is
natural to study the adjoint  process and its limiting behavior is
\begin{multline}
\mathcal Y^+_{x_0}=\Big\{\bigl\{x(t),\chi(t)\bigr\}_{t\in [0,T]} \in
C([0,T], \Omega)\times\mathcal M_0([0,T])^{\Gamma}:\\
\sum_{\sigma\in \Gamma}\chi_{\sigma}(t)=1\ a.e.,\
x(t)=x_0-\int_0^t\sum_{\sigma\in
\Gamma}\chi_\sigma(s)F_{\sigma}(x(s))\,ds\Big\}\,. \label{spazio+}
\end{multline}
For models having invariant measure of the form \eqref{guess}, the
rates $r^+$ of the adjoint process do not depend on $\lambda$ so
that a LDP for the adjoint process can be obtained  using again the
results of \cite{FGRmat}:
\begin{equation}\label{marchioro}
\mathbb
P^{\lambda,+}_{x_0,\sigma_0}\left(\bigl\{x(t),\chi(t)\bigr\}_{t\in
[0,T]}\approx \left\{\hat{x}(t),\hat{\chi}(t)\right\}_{t\in
[0,T]}\right)\sim e^{-\lambda
J_{[0,T]}^+\bigl(\bigl\{\hat{x}(t),\hat{\chi}(t)\bigr\}_{t\in[0,T]}\bigr)}\,,
\end{equation}
where
\begin{align}
& J^+_{[0,T]}\bigl(\bigl\{x(t),\chi(t)\bigr\}_{t\in[0,T]}\bigr):=
\int _0^T j^+ \bigl ( x (t) ,\chi (t)  \bigr) dt\,,\nonumber \\
& j^+(x, \chi ):=\sup _{z\in (0,\infty)^\G} \sum _{(\s,\s')\in W }
\chi_\s  r^+(\s,\s'|x)\left [ 1-\frac{z_{\s'}}{z_\s } \right]
\,\label{pranzo?}.
\end{align}
 In the above formula $\bbP^{\lambda,+}_{x_0,\sigma_0}$ is the
probability measure on $\mathcal Y^+_{x_0}$ induced by the adjoint
process with parameter $\lambda$ and initial condition
$(x_0,\sigma_0)$. Remember that in this case the rates $r^+$ in
\eqref{pranzo?} are related to the rates $r$ of the direct model
from \eqref{r+}.

Let us assume now, as done for \eqref{density}, that  for all $x \in
\O$ the chemical part $L_c[x]$ of the direct generator is reversible
w.r.t. the quasistationary measure $\mu(\cdot |x)$. Then we know
that the same property holds for the adjoint process with
$\mu(\cdot|x)$ replaced by $\mu ^+(\cdot|x)$. In this case,
similarly to \eqref{density}, we get
\begin{equation}
j^+(x,\chi)=\sum_\sigma \g^+(\sigma|x)\chi_\sigma- \sum_{(\s,\s')\in
W}
\sqrt{\frac{\mu^+(\sigma|x)}{\mu^+(\sigma'|x)}}r^+(\sigma,\sigma'|x)
\sqrt{\chi_\sigma}\sqrt{\chi_{\sigma'}}\,.
\end{equation}
Recalling relations \eqref{r+} and \eqref{quaqua+=invnoL} we obtain
that
\begin{equation}
j^+(x,\chi) =  \sum_\sigma  \gamma^+ (\s|x)\chi_\s -  \sum
_{(\s,\s')\in W}
\sqrt{\frac{\mu(\sigma|x)}{\mu(\sigma'|x)}}r(\sigma,\sigma'|x)
\sqrt{\chi_\sigma}\sqrt{\chi_{\sigma'}}\,.\label{density?}
\end{equation}

   As the reader can check, the proof of the LDP in \cite{FGRmat}
remains valid for PDMPs with $\l$--dependent  rates $r
(\s,\s',\l|x)$ obtained as perturbation of $\l$--independent rates,
i.e. $r(\s,\s',\l|x) = r(\s,\s'|x)(1+o(1))$. Hence, the above result
\eqref{marchioro} can be extended   to more general   processes with
invariant measures not of the form \eqref{guess}. We will discuss an
example in the next section.

\medskip

Let us now consider  the LD rate functional $J^{m,+}_{[0,T]}:
C([0,T], \O) \rightarrow [0,\infty]$ for the evolution of the
mechanical state in the adjoint process (dropping the above the
reversibility assumption). It
 has
the form
\begin{equation}
J_{[0,T]}^{m,+}\left(\{x(t)\}_{t\in [0,T]}\right)= \begin{cases}
\int_0^Tj^+_m (x(t),\dot{x}(t))\,dt  & \text{ if } x(\cdot) \in
\cY_{x_0} ^{m,+} \,,\\ +\infty & \text{ otherwise}\,,
\end{cases}
\end{equation}
where, similarly to $\cY_{x_0}^m$, the space $\cY_{x_0} ^{m,+}$ is
defined as the mechanical projection of $\cY^+_{x_0}$, while the the
density $j^+_m(x,\dot{x})$ is given by
\begin{equation}
j_m^+ (x,\dot{x})=\inf_{\left\{\chi:\ \dot{x}=-\sum_\sigma
\chi_\sigma F_{\sigma}(x)\right\}}j^+(x,\chi)\,.\label{densitym+}
\end{equation}
In the above formula, $\chi$ varies among the vectors in $[0,1]^\G$
such that $\sum_\s \chi_\s =1$.

If the collection of vectors $\left\{F_\s(x)\right\}_{\s\in \Gamma}$
are the vertices of a simplex for any $x\in \Omega$ then this holds
also for the vectors $\left\{-F_\s(x)\right\}_{\s\in \Gamma}$ and
consequently we have
\begin{equation}
j_m^+(x,\dot{x})=
\begin{cases}
j(x,\chi^{-F}(\dot{x})) & \text{ if } \dot{x}\in \mathcal
C\left(\left\{-F_\sigma(x)\right\}_{\sigma\in \Gamma}\right)\,,\\
+\infty & \text{ otherwise}\,.
\end{cases}
\label{densitymsimplex+}
\end{equation}
Trivially, $ \cC\bigl( \{ -F_\s(x) \}_{\s \in \G } \bigl)=-\cC\bigl(
\{ F_\s(x) \}_{\s \in \G } \bigl)$ and $ \chi ^{-F} ( \dot{x} ) =
\chi ^F (- \dot{x} ) $.

\section{Fluctuation theory}\label{nevissima}
In this section we further investigate  the fluctuations of the
mechanical variables of our  PDMPs, following ideas and results
developed in \cite{BDGJL1, BDGJL2, BDGJL3, BDGJL4} for  interacting
particle systems and inspired by the Freidlin and Wentzell theory
\cite{FW} for diffusion processes.
 As we will  show, PDMPs  are a natural
source of examples where  the {\it macroscopic fluctuation theory}
developed in \cite{BDGJL1, BDGJL2, BDGJL3, BDGJL4} applies. A key
identity in this theory is given by the Fluctuation--Dissipation
(FD) relation \eqref{quintettsim}, which in \cite{BDGJL2} is a
direct consequence of the Markov property, while the mechanical
evolution of our PDMPs is  not Markov. Hence, the FD relation cannot
be taken for granted in our case. In Subsection \ref{pppp} we will
prove it for the class of exactly solvable PDMPs with stationary
measure given by \eqref{guess} as well for PDMPs on the 1d torus not
satisfying \eqref{guess}. Before considering these cases, in
Subsection \ref{degregori} we reformulate the results of
\cite{BDGJL1, BDGJL2, BDGJL3, BDGJL4}, in the simpler context of
processes with trajectories in $C([0,T],\O)$ and discuss
consequences of the FD relation \eqref{quintettsim}.

\subsection{General framework}\label{degregori}

We consider a  $\l$--parameterized family of stochastic Markov
processes with trajectories in $C([0,T],\O)$, satisfying a sample
path LD principle  as the parameter $\lambda$ diverges to $+\infty$.
This means that, fixed $\{\hat x (t) \}_{t \in [0,T]}\in
C([0,T],\O)$, it holds
\begin{equation}
\mathbb P_{x_0}^\lambda\left(\{x(t)\}_{t\in[0,T]}\approx\left\{\hat
x(t)\right\}_ {t\in[0,T]}\right)\sim e^{-\lambda
I_{[0,T]}^{x_0}\left(\{\hat x(t) \}_{t\in[0,T]}\right)}\,,
\label{LDPgen}
\end{equation}
where in  the above formula $\{x(t) \}_ {t\in[0,T]}$ denotes a
generic element of $C([0,T],\O)$ and $\mathbb P_{x_0}^\lambda$
denotes  the law on $C([0,T],\O)$ induced by the $\l$--parameterized
process with  initial configuration
 $x_0\in \O$. As a prototype one can take diffusions on
 $\O=\mathbb R^d$ with noise of order $\sqrt{1/\l}$ as in the Freidlin and Wentzell
 theory \cite{FW}.

We further assume that  for any fixed $\lambda$ the
$\l$--parameterized process admits  a unique invariant measure
$\rho_\lambda$. Then the adjoint process can be defined and has
$\rho_\l$ as unique invariant measure. We assume that also the
$\l$--parameterized family of adjoint processes satisfies a LD
principle  as $\l$ diverges, i.e. \eqref{LDPgen} remains valid with
$\bbP ^\l_{x_0}$ and $I^{x_0}_{[0,T]}$ replaced by
 $\bbP
^{\l,+}_{x_0}$ and $I^{x_0,+}_{[0,T]}$, respectively. In addition,
we assume that there exist densities $\cL(x,\dot x), \cL^+(x, \dot
x): \O \times \bbR^d \rightarrow [0,\infty)$ such that for any
initial configuration $x_0$  the rate functionals  $I^{x_0}_{[0,T]}$
and $I^{x_0,+}_{[0,T]}$ admit an  integral representation of the
form
\begin{align}
& I_{[0,T]}^{x_0}\left(\{x(t)\}_{t\in[0,T]}\right)=
\begin{cases}
\int_0^T\mathcal L(x(t),\dot{x}(t))dt & \text{ if } x(\cdot) \in  \cX_{x_0}\,,\\
+\infty & \text{ otherwise}\,,\\
\label{din1}
\end{cases}
\\
& I_{[0,T]}^{x_0,+}\left(\{x(t)\}_{t\in[0,T]}\right)=
\begin{cases}
\int_0^T\mathcal L^+(x(t),\dot{x}(t))dt & \text{  if }
x(\cdot)=\cX_{x_0}^+ \,,\\
+\infty & \text{ otherwise}\,, \label{din2}
\end{cases}
\end{align}
for suitable subspaces $\cX_{x_0}, \cX^+_{x_0}\subset C([0,T], \O)$.
This assumption implies in particular that all paths in $\cX_{x_0}$
and $\cX_{x_0}^+$ starts in $x_0$.

 The functions $\mathcal L$ and $\mathcal L^+$ are called
Lagrangians. Typically,  $\cL (x, \dot x )=0$ if and only if $\dot
x=\bar{F}(x)$ for a suitable vector field $\bar{F}(x)$ that
identifies  the law of large numbers of the model in the limit of
diverging $\lambda$. In fact, in this case, from \eqref{LDPgen} we
can immediately derive that
\begin{equation}\label{susi}
\lim_{\lambda \to +\infty}\mathbb
P^{\lambda}_{x_0}\Big(\sup_{t\in[0,T]}|x(t)-\bar{x}(t)|>\delta\Big)=0\,,
\ \ \ \ \forall \delta>0\,,
\end{equation}
where $\left\{\bar{x}(t)\right\}_{t\in[0,T]}\in C([0,T],\O)$ solves
the Cauchy problem
\begin{equation}
\left\{
\begin{array}{l}
\dot{\bar{x}}(t)=\bar{F}(\bar{x}(t))\,,\\
\bar{x}(0)=x_0\,.\\
\end{array}
\right.\label{eqdiffdir}
\end{equation}
Similarly we require that $\mathcal L^+$  vanishes along a path
$\{x(t)\}_{t \in [0,T]}$  if and only if $\dot
x(t)=\bar{F}^+(x(t))$, where  the vector field $\bar{F}^+$
identifies the law of large numbers of the adjoint process in the
limit of diverging $\lambda$.

Finally we assume that the family of invariant measures
$\rho_\lambda$ satisfy a LD principle on $\O$ as $\l$ diverges, with
rate function $V(x)$. This means that for any fixed $t$ and for any
fixed $x\in \O$ it holds
\begin{equation}\label{kreacher}
\mathbb P_{\rho_\l }^\lambda\left(x(t)\approx x\right)\sim
e^{-\lambda V(x)}\,,
\end{equation}
where   $\mathbb P_{\rho_\l}^\lambda$  denotes the law of the
stationary process with parameter $\l$ and  initial distribution
$\rho_\l$. Below we will denote by $\mathbb P_{\rho_\l}^{\lambda,+}$
the law of its adjoint process.

\medskip

Let us now derive some consequences from our assumptions.  By
definition we have for any  path $x(\cdot) \in C([-T,T], \O)$ that
\begin{equation}\label{alba} \mathbb P_{\rho_\l
}^\lambda\left(\left\{X(t)\right\}_{t\in[-T,T]}\approx
\left\{x(t)\right\}_{ t\in [-T,T]}\right)=\mathbb
P_{\rho_\l}^{\lambda,+}\left(\left\{X(t)\right\}_{t\in
[-T,T]}\approx \left\{x(-t)\right\}_{t\in [-T,T]}\right)\,.
\end{equation}
  Due to the fact that the processes are Markov and that the path
$\{x (t-T) \} _{t\in [0,2T]}$ belongs to the path space $\cX_{
x(-T)}$ (referred to the time interval $[0,2T])$, while the path $\{
x(T-t)\}_{t \in [0,2T]} $ belongs to the path space $\cX _{x (T) }$
(referred to the time interval $[0,2T])$, \eqref{alba}
  implies the following relation concerning the LD rate functionals:
\begin{equation}
V(x(-T))+\int_{-T}^T\mathcal
L(x(t),\dot{x}(t))\,dt=V(x(T))+\int_{-T}^T\mathcal
L^+(x(-t),-\dot{x}(-t))\,dt\,. \label{quintett}
\end{equation}
Dividing both sides of \eqref{quintett} by $2T$ and taking the limit
$T\downarrow 0$ we obtain FD relation
\begin{equation}
\mathcal L(x,\dot{x})=\nabla V(x)\cdot \dot{x}+\mathcal
L^+(x,-\dot{x})\,, \label{quintettsim}
\end{equation}
 valid for any $x,\dot x$ corresponding to the values  $x(0)$, $\dot x(0)$ for some path $x(\cdot )$ as above.

\smallskip
From now on we suppose that the FD relation \eqref{quintettsim}
holds, without assuming that the processes under consideration are
Markov. Following \cite{BDGJL2} we derive some consequences of
\eqref{quintettsim}.

\smallskip
A point $x_*\in \O$ is  called  {\it equilibrium point} for the
$\l$--parameterized family of processes  if
 $\bar{F}(x_*)=0$. Then,   by the LLN  \eqref{susi},  the trajectory $\bar x(t)\equiv x_*$
 is the limiting path for the process starting in $x_*$, thus implying that
 $I^{x_*}_{[0,T]} (\bar x(\cdot ) )=0$, i.e.  $\mathcal L(x_*,0)=0$.
 By means of
the FD relation  we obtain that  $\mathcal L^+(x_*,0)=0$ and
consequently $x_*$ is an equilibrium point also for the family of
adjoint processes, i.e. we have $\bar{F}^+(x_*)=0$. We restrict now
to the case that the vector field $\bar F$ has  a unique equilibrium
point $x_*$, which is a global attractor. This means that
$$
\lim_{t\to +\infty}x(t)=x_*\,
$$
for any $\left\{x(t)\right\}_{t\in[0,+\infty)}$ solving
$\dot{x}=\bar{F}(x)$.  Note that, due to the previous  argument,
$x_*$ is also the unique equilibrium point of the vector field
$\bar{F}^+$. We assume that it is also a global attractor for $\bar
F^+$. As  simple example satisfying all the above assumptions,
consider   the family of reversible diffusions on $\O=\mathbb R^d$
descried by the SDE
$$ \dot x= -\nabla U(x)+\lambda^{-\frac 12}\dot
w\,,
$$
where $U$ is a single well potential and $w$ is a standard Brownian
motion. In this case
$$
\mathcal L(x,\dot x)=\mathcal L^+(x,\dot x)=\frac 12 \left|\dot
x+\nabla U(x)\right|^2\,,
$$
and $V(x)=2U(x)$.

Let us introduce  the {\it quasi-potentia}l $Q(x)$ as function on
$\O$ defined by
\begin{equation}
Q(x)=\inf_{\mathcal A_x} \int _{-\infty} ^0 \cL ( x(t), \dot x (t)
)dt\,, \label{quasipot}
\end{equation}
where
$$
\mathcal A_x=\Big\{\left\{x(t)\right\}_{t\in (-\infty,0]}\ : \
x(0)=x\ ,\ \lim_{t\to -\infty}x(t)=x_*\Big\}\,.
$$
In the case  of multiple equilibrium points and different basins of
attraction the definition has to be suitably  modified. We will not
consider this situation here, referring to \cite{FW} the interested
reader.

\begin{Rem}
We point out that the rigorous  definition of  quasi--potential is
slightly different from \eqref{quasipot}. Indeed, since  the dynamic
LD principles with rate functionals \eqref{din1}, \eqref{din2} hold
for finite time intervals, one  has to   define the quasi--potential
as
\begin{equation}
Q(x)=\inf_{T\geq 0}\inf_{\mathcal A_x^T} \int _{-T} ^0 \cL ( x(t),
\dot x (t) )\, dt\,, \label{trueQ}
\end{equation}
where
$$
\mathcal A_x^T=\Big\{\left\{x(t)\right\}_{t\in [-T,0]}\ : \ x(0)=x\
,\ x(-T)=x_*\Big\}\,.
$$
We use definition \eqref{quasipot} to simplify our discussion, while
the interested reader can adapt our arguments in order to obtain
rigorous proofs (see for example \cite{BDGJL5}).
\end{Rem}

\smallskip
 We now derive an {\it $H$--Theorem} for the quasi--potential. More
 precisely, we show that
the quasi--potential is a decreasing Lyapunov functional  for both
the vector field $\bar{F}$ and for the vector field $\bar{F}^+$.
This means that $t\rightarrow Q (x(t))$ is a decreasing  function if
$\dot x(t)= \bar F (x(t))$ or if
 $\dot x(t)= \bar F^+ (x(t))$, respectively.
In order to justify our claim, we take a path $\{x(t)\}_{t\geq 0 }$
such that $\dot x(t)= \bar F (x(t))$. Let $x=x(0)$ and $x'=x(t')$,
with $t'>0$. Given $\e>0$ we fix an element
$\left\{\widetilde{x}(t)\right\}_{t\in[-\infty,0]}\in \mathcal A_x$
such that $Q(x)$ differs from $I_{(-\infty,0] } (\tilde x (\cdot) )$
at  most $\e$.  Then we construct the following element of $\mathcal
A_{x'}$
\begin{equation*}
\bar{x}(t)=
\begin{cases}
\widetilde{x}(t+t') & \text{ if } \ t\leq -t'\,,\\
 x(t+t') & \text{ if }   t\in (-t',0]\,.
\end{cases}
\end{equation*}
Since $\cL(x(t),\dot x(t))=0$ for all $t\geq 0$,  we  have
\begin{multline}  Q(x')\leq  \int_{-\infty}^0\mathcal L(\bar x (t),\dot{\bar{x}} (t))\,dt=
 \int_{-\infty}^{-t'}\mathcal
L(\widetilde{x}(t+t'),\dot{\widetilde{x}}(t+t'))\,dt+\\\int_{-t'}^0\mathcal
L(x(t+t'),\dot x(t+t'))\,dt= \int_{-\infty}^0\mathcal
L(\widetilde{x}(t),\dot{\widetilde{x}}(t))\,dt\leq Q(x)+\e
\,.\nonumber
\end{multline}
By the arbitrariness of $\e$, we deduce that   $Q(x')\leq Q(x)$.
Therefore, it must be   $\nabla Q(x)\cdot \bar{F}(x)\leq 0$ for all
$x \in \O$. The same kind of argument can be used for the vector
fields $\bar{F}^+$, thus implying that  $\nabla Q(x)\cdot
\bar{F}^+(x)\leq 0$.

\smallskip

Let us now show that the quasi--potential  $Q(x)$ coincides with the
LD rate functional $V(x)$  of the invariant measures $\rho_\l$:
$$
Q(x)=V(x) \qquad \forall x \in \O\,.
 $$
Since  $\mathcal L^+$ is nonnegative, using the FD relation  we get
for any $\left\{x(t)\right\}_{t\in [-\infty,0]}\in \mathcal A_x$
that
\begin{equation}\label{cri}
\int_{-\infty}^0\mathcal L(x(t),\dot{x}(t))\,dt\geq
\int_{-\infty}^0\nabla V(x(t))\cdot \dot{x}(t)\,dt
=V(x)-V(x_*)=V(x)\,.
\end{equation}
The last equality follows from the fact that the rate functional $V$
is zero on the unique equilibrium  point $x_*$. The function $V$ is
in fact non negative due to the fact that it is a rate functional
and it is zero in correspondence of $x^*$ that corresponds to the
typical value (law of large numbers). Due to the definition of the
quasi--potential $Q$, the above bound \eqref{cri} implies that $Q(x)
\geq V(x)$. In order to prove the reversed inequality, let
$\left\{x^+(t)\right\}_{t\in [0,+\infty]}$ be the solution of the
Chauchy problem
\begin{equation}
\left\{
\begin{array}{l}
\dot{x}^+(t)=\bar{F}^+(x^+(t))\,,\\
x^+(0)=x\,.\\
\end{array}
\right.\label{eqdiffadj}
\end{equation}
Due to the  global attractiveness of $x_*$ we have
$$
\lim_{t\to +\infty}x^+(t)=x_*\,,
$$
so that $\mathcal T \Big[\left\{x^+(t)\right\}_{t\in
[0,+\infty]}\Big]=\left\{x^+(-t)\right\}_{t\in [-\infty,0]}\in
\mathcal A_x$ and, by definition of $Q(x)$ and due to the FD
relation \eqref{quintettsim},
$$
Q(x) \leq \int_{-\infty}^0\mathcal
L(x^+(-t),-\dot{x}^+(-t))\,dt=V(x)\,,
$$
thus concluding the proof  that $Q(x)=V(x)$. Coming back to the
above expression, we then conclude that the path
$\left\{x^+(-t)\right\}_{t\in [-\infty,0]}$ is the minimizer in
\eqref{quasipot}.
 Hence, we arrive at the following key observation. Starting from
 equilibrium at time zero, for $\l$ and $T$ large, if the system at time $T$ is
 in  state $x$ then with high probability its evolution for times $t\in [0,T]$ is well approximated by the path $x^+( T-\cdot)$, where
$x^+(\cdot)$ solves
 \eqref{eqdiffadj}. More precisely:
\begin{multline}
\lim _{T\uparrow \infty}\lim _{\l \uparrow \infty} \bbP^\l_{\rho_\l}
\left( \{x (t) \}_{t \in [0,T]} \approx  \{ x^+ (T-t) \}
_{t \in [0,T]}  \Big| x(T)=x \right) =\\
\lim _{T\uparrow \infty}\lim _{\l \uparrow \infty} \bbP^\l_{\rho_\l}
\left( \{x (t) \}_{t \in [-T,0]} \approx  \{ x^+ (-t) \} _{t \in
[-T,0]} \Big| x(0)=x \right)= 1\,.
\end{multline}
We call $\{x^+(-t)\}_{t \in [0,\infty]}$ the {\sl exit trajectory},
while we call  the path $\{\bar{x}(t)\}_{t\in [0,+\infty]}$ solving
\eqref{eqdiffdir} the {\sl relax trajectory} (motivated by the LLN).
When the vector fields $\bar{F}$ and $\bar{F}^+$ coincide, i.e.
\begin{equation}
\bar{F}(x)=\bar{F}^+(x)\,, \ \ \ \ \ \ \ \forall x\in
\O\,,\label{+-}
\end{equation}
then the exit/relax trajectories are related by time reversal and
using the terminology of \cite{BDGJL1}, \cite{BDGJL2} we say that an
{\it Onsager-Machlup symmetry} holds. When condition \eqref{+-} does
not hold, the exit/relax trajectories are not necessarily related by
time reversal and according to \cite{BDGJL1}, \cite{BDGJL2} we say
that a {\it generalized Onsager-Machlup symmetry} holds.

\smallskip

We conclude this subsection justifying the name ``Lagrangian" given
to $\cL(x,\dot x)$.
 From classical arguments in
variational analysis it follows that the quasi--potential $Q(x)$ as
defined in \eqref{quasipot} solves the Hamilton-Jacobi equation
\begin{equation}
\mathcal H(x,\nabla Q(x))=0\,, \label{HJ}
\end{equation}
where the Hamiltonian $\mathcal H$ is obtained as Legendre transform
of $\mathcal L$ as
\begin{equation}
\mathcal H(x,p)=\sup_{y\in \bbR^d}\left(p\cdot y-\mathcal L
(x,y)\right)\,.\label{Ham}
\end{equation}
It can be shown   (see \cite{BDGJL2} for details) that $Q$ is the
maximal solution of \eqref{HJ}.  The r.h.s. of \eqref{HJ} must be
zero for the following reason. Due to \eqref{Ham} it must be
$\cH(x_*, 0)=0$, while due to the fact that $Q(x) \geq Q(x_*)$ it
must be $\nabla Q(x_*)=0$. Hence, $\cH(x_*, \nabla Q(x_*))=0$.

Clearly the above arguments hold also for the  family of adjoint
processes. In particular,  the quasi--potential $Q$  solves also
Hamilton-Jacobi equation
\begin{equation}
\mathcal H^+(x,\nabla Q(x))=0\,, \label{HJ+}
\end{equation}
where the Hamiltonian $\mathcal H^+$ is obtained as Legendre
transform of $\mathcal L^+$ as
\begin{equation}
\mathcal H^+(x,p)=\sup_{y\in \mathbb R^d}\left(p\cdot y-\mathcal L^+
(x,y)\right)\,.\label{Ham+}
\end{equation}
Note that due to the validity of the FD relation \eqref{quintettsim}
we have
\begin{eqnarray*}
& & \mathcal H(x,p)=\sup_{y\in \mathbb R^d}\left(p\cdot y-\mathcal
L (x,y)\right)\\
& &= \sup_{y\in \mathbb R^d}\left(p\cdot y -\nabla V(x)\cdot
y-\mathcal L^+ (x,-y)\right)\\
& &=\sup_{y\in \mathbb R^d}\left((\nabla V(x)-p)\cdot y-\mathcal L^+
(x,y)\right)=\mathcal H^+(x,\nabla V(x)-p)\,.
\end{eqnarray*}

\subsection{Application to PDMPs}\label{pppp} In this subsection we
further analyze the fluctuations of  the process $x(t)$ describing
the evolution of the mechanical state in our PDMPs, according to the
results just described. As already stressed, the theory  described
in the previous subsection is based on the FD relation
\eqref{quintettsim}, which is always true if the process under
consideration is Markov. In the case of PDMPs, the Markov property
of $x(t)$ is typically violated. We check here the validity of the
FD relation for a large class of PDMPs. That automatically implies
the  discussion following \eqref{quintettsim}.

\subsubsection{PDMPs with invariant measure of the form
\eqref{guess}} Let us first restrict to PDMPs with invariant measure
of the form \eqref{guess}, for which we can apply the results of
Section \ref{HFLLD} also to the adjoint process. The
 vector fields
$\bar{F}$ and $\bar{F}^+$ mentioned in the previous subsection
become
\begin{equation}
\left\{
\begin{array}{l}
\bar{F}(x)=\sum_\sigma \mu(\sigma|x)F_\sigma(x)\,,\\
\bar{F}^+(x)=-\sum_\sigma \mu^+(\sigma|x)F_\sigma(x)\,.\\
\end{array}
\right.\nonumber
\end{equation}
We recall that  $\mu(\cdot |x)$ and $\mu^+(\cdot |x)$ are the unique
invariant measures of $L_c[x]$ and $L^+_c [x]$, respectively.

We already stressed that the mechanical process is in general not
Markovian and consequently there is not a natural notion of
invariant measure. This has to be replaced by the projection $\hat
\rho_\lambda $ on the $x$ component of the invariant measure of the
joint mechanical and chemical Markov process
$$
\hat \rho_\l (x):=\sum_\sigma \rho_\lambda(x,\sigma)\,.
$$
In this case it is easy to compute the limit $\lim _{\l\to \infty}
\l^{-1} \log \hat \rho_\l (x)$, and derive a LD principle for $\hat
\rho$. To this aim, first  observe that the function $S$ in
\eqref{guess} is univocally determined up to an additive constant.
From now on, we denote by  $S$ the unique function satisfying
\eqref{guess}  normalized in such a way that
%\begin{equation}
$\inf _{x \in \O} S(x) =0$.
%\end{equation}
By this choice, it is simple to see that
$$
\lim_{\lambda \to \infty}\lambda^{-1}\log \hat \rho _\l(x)=-S(x)\,.
$$
Comparing with \eqref{kreacher}, we deduce that $S(x)=V(x)$, namely
the function $S$ coincides with the LD rate functional of the
measures $\hat \rho_\l$.

We now establish for this class of models the symmetry relation
\begin{equation}\label{cruciale}
j(x,\chi)- j^+(x,\chi)= \sum _\s \chi_\s\left( \g (\s|x) - \g
^+(\s|x) \right)\,,
\end{equation}
for the joint rate density $j$ and $j^+$ introduced in \ref{HFLLD}.

When $L_c [x]$ and consequently also $L_c^+[x]$ are reversible
w.r.t. the corresponding quasistationary measures, then
\eqref{cruciale} follows directly from the explicit expressions
\eqref{density} and \eqref{density?}. In the general case we can
write the variational expressions \eqref{pranzo} and \eqref{pranzo?}
as
\begin{align}
&j(x,\chi)=\sum_\s \chi_\s \gamma(\s|x)-\inf_{z\in
(0,+\infty)^\Gamma}\left(\sum_{(\s,\s')\in
W}\chi_\s r(\s,\s'|x)\frac{z_{\s'}}{z_\s}\right)\,,\label{1v}\\
&j^+(x,\chi)=\sum_\s \chi_\s \gamma^+(\s|x)-\inf_{z\in
(0,+\infty)^\Gamma}\left(\sum_{(\s,\s')\in
W}\chi_{\s'}r(\s,\s'|x)\frac{\rho(x,\s)z_{\s}}{\rho(x,\s')z_{\s'}}\right)\,.\label{2v}
\end{align}
When $\chi_\s>0$ for any $\s\in \Gamma$ we introduce
$\widetilde{z}_\s:=\frac{\chi_\s}{\rho(x,\s)z_\s}$. The variational
expression in \eqref{2v} can then be written as
$$
\inf_{\widetilde{z}\in (0,+\infty)^\Gamma}\left(\sum_{(\s,\s')\in
W}\chi_\s
r(\s,\s'|x)\frac{\widetilde{z}_{\s'}}{\widetilde{z}_\s}\right)\,,
$$
that coincides with the variational expression in \eqref{1v}.
Relation \eqref{cruciale} now follows immediately. The same result
can be obtained also in the case that  there exists some $\s\in \G$
for which $\chi_\s=0$. We discuss this case in Appendix
\ref{apemaia}.

Using equations \eqref{eqeq} (valid for each $\s \in \G)$, which are
equivalent to the second group of equations in \eqref{stazguess}, we
obtain that \eqref{cruciale} can be written as
\begin{equation}\label{cruciale2}
j(x,\chi)- j^+(x,\chi)= \nabla S(x)\cdot \left(\sum _\s \chi_\s
F_\s\right)\,.
\end{equation}
Observe that to compute both $j_m(x,\dot x)$ and $j_m^+(x,-\dot x)$
we need to minimize respectively $j$ and $j^+$ over $\chi$ subject
to the same constraint
\begin{equation}
\left\{\sum_s \chi_s F_\s(x)=\dot x\right\}\,. \label{constraint}
\end{equation}
Recalling \eqref{cruciale2}, we get  for any fixed $x$ and $\dot x$
that
\begin{eqnarray}
j_m(x,\dot x)&=&\inf_{\left\{\chi:\ \sum_s \chi_s F_\s(x)=\dot
x\right\}}j(x,\chi)\nonumber \\
&=&\nabla S(x)\cdot \dot x +\inf_{\left\{\chi:\
\sum_s \chi_s F_\s(x)=\dot x\right\}}j^+(x,\chi) \nonumber \\
&=&\nabla S(x)\cdot \dot x+j_m^+(x,-\dot x)\,. \label{kkk}
\end{eqnarray}
That corresponds to the FD relation \eqref{quintettsim} since, as
already observed, $S(x)=V(x)$. From \eqref{kkk} we obtain also that
the minimizers in \eqref{constraint} for the computation of both
$j_m(x,\dot x)$ and $j_m^+(x,-\dot x)$, coincide. We point out
that in the case of computable LD rate functionals as in
\eqref{densitymsimplex} and \eqref{densitymsimplex+}, identity
\eqref{kkk} follows directly from \eqref{density}, \eqref{density?}
and the simple relation $\chi_\s^F(\dot x)=\chi_\s^{-F}(-\dot x)$.

\smallskip

It is interesting to note that we can obtain a simple explicit
expression for the {\it entropy production}, i.e.  the rate of
variation of $S$ along the orbits of the vector fields $\bar F$ and
$\bar{F}^+$. Given $x(t)$ a solution of \eqref{eqdiffdir}, using
\eqref{eqeq} we have
\begin{equation}
\nabla S(x)\cdot \dot{x}=\nabla S(x)\cdot \bar{F}(x)=\sum_{\s}\mu
(\s|x)\nabla S(x)\cdot F_\s(x) = \sum_{\s}\mu
(\s|\bar{x})\left(\gamma(\s|x)-\gamma^+(\s|x)\right)\,.
\label{entpro}
\end{equation}
Likewise given $x^+$ a solution of \eqref{eqdiffadj} we have
\begin{equation}
\nabla S(x^+)\cdot \dot{x}^+=\sum_{\s}\mu^+
(\s|x^+)\left(\gamma^+(\s|x^+)-\gamma(\s|x^+)\right)\,.
\label{entrpro+}
\end{equation}

\bigskip

We now illustrate with examples the validity of the general results
discussed in subsection \ref{degregori} for PDMPs. We stress that
these results are a direct consequence  of the FD relation
\eqref{quintettsim} checked above. We first consider the 1D models
of Subsection \ref{1dbound}. In this case the quasistationary
measures for the direct and adjoint chemical generators can be
easily computed and we have
\begin{equation}
\begin{cases}
\bar{F}(x)= \frac{r(1,0|x)}{r(1,0|x)+r(0,1|x)}F_0(x)+\frac{r(0,1|x)}{r(1,0|x)+r(0,1|x)}F_1(x)\,,\\
\bar{F}^+(x)=-\left(\frac{r(1,0|x)F_0^2(x)}{r(1,0|x)F_0^2(x)+r(0,1|x)F_1^2(x)}F_1(x)+
\frac{r(0,1|x)F_1^2(x)}{r(1,0|x)F_0^2(x)+r(0,1|x)F_1^2(x)}F_0(x)\right)\,.\\
\end{cases}
\end{equation}
Note that   $x \in \O$ is an equilibrium point for $\bar F$ if and
only if
$$
r(1,0|x)F_0(x)+r(0,1|x)F_1(x)=0.
$$
The same equation characterizes the equilibrium points of $\bar
F^+$, thus implying that $\bar F$ and $\bar F^+$ have the same
equilibrium points in $\O$. Moreover we have that both relations
$\bar{F}(x)>0$ and $\bar{F}^+(x)>0$ holds if and only if
$$
r(1,0|x)F_0(x)+r(0,1|x)F_1(x)>0,
$$
so that also the stability of the equilibrium points for both vector
fields is the same. We point out that, since $F_0(a)=F_1(b)=0$,
 the  adjoint vector field $\bar{F}^+$ has two
additional equilibrium points at the boundary $\partial
\Omega=\{a,b\}$, which are necessary unstable.

Note that there are models such that the vector field $\bar{F}$ has
many stable equilibrium points. In this case the definition of the
quasi--potential $Q(x)$ in \eqref{quasipot} has to be modified
considering separately the different basin of attraction. We do not
discuss this possibility.

For a vector $\dot{x}\in \mathcal C\left\{F_0(x),F_1(x)\right\}$ we
have that
$$
\chi_0^F(\dot{x})=\frac{F_1(x)-\dot{x}}{F_1(x)-F_0(x)}\,, \ \ \ \ \
\ \ \ \chi_1^F(\dot{x})=\frac{\dot{x}-F_0(x)}{F_1(x)-F_0(x)}\,.
$$
Using the results of section \ref{HFLLD} (see \eqref{due} and
\eqref{densitymsimplex}), we get
\begin{equation}
j_m(x,\dot{x})=\left(\sqrt{\frac{r(0,1|x)(F_1(x)-\dot{x})}{F_1(x)-F_0(x)}}
-\sqrt{\frac{r(1,0|x)(\dot{x}-F_0(x))}{F_1(x)-F_0(x)}}\right)^2
\label{quellagia}
\end{equation}
for all $x \in \O$ and $\dot x \in \cC (F_0(x), F_1(x) )$. Since
similarly to \eqref{due}+\eqref{densitymsimplex} it holds
\begin{equation}\label{sissi} j_m^+(x,\dot x) = \left( \sqrt{ \chi^{-F}_0(\dot x)
r^+(0,1|x) } -\sqrt{ \chi^{-F} _1 (\dot x) r^+ (1,0|x) }
\right)^2\,,
\end{equation}
observing that $\chi ^{-F}(\dot x)= \chi ^F(-\dot x)$ for all $\dot
x \in \cC (-F_0(x),-F_1(x) )$ we conclude that
\begin{equation}\label{carletto}
j_m^+(x,\dot{x})=\left(\sqrt{\frac{r(1,0|x)|F_0(x)|(F_1(x)+\dot{x})}{F_1(x)(F_1(x)-F_0(x))}}
-\sqrt{\frac{r(0,1|x)F_1(x)(-\dot{x}-F_0(x))}{|F_0(x)|(F_1(x)-F_0(x))}}\right)^2
\end{equation}
for all $x \in \O$ and $\dot x \in \cC (-F_0(x),-F_1 (x) )$. The
validity of the FD relation \eqref{quintettsim} can now be checked
directly recalling that in this case
\begin{equation}
\nabla S(x)=\frac{r(0,1|x)}{F_0(x)}+ \frac{r(1,0|x)}{F_1(x)}\,.
\label{S1d}
\end{equation}
Also the validity of the H--Theorem can be checked directly, we have
in fact
$$
\nabla S(x)\cdot
\bar{F}(x)=-\frac{\left(r(1,0|x)\sqrt{\frac{|F_0(x)|}{F_1(x)}}
+r(0,1|x)\sqrt{\frac{F_1(x)}{|F_0(x)|}}\right)^2}{r(0,1|x)+r(1,0|x)}\,,
$$
and
$$
\nabla S(x)\cdot
\bar{F}^+(x)=-\frac{\Big(r(0,1|x)F_1(x)+r(1,0|x)F_0(x)\Big)^2}{r(0,1|x)F_1^2(x)+r(1,0|x)F_0^2(x)}\,,
$$
whose negativity is immediate. Finally we can also explicitly
compute the Hamiltonian
$$
\mathcal H (x,p)=\sup_{y\in \mathcal
C\bigl\{F_0(x),F_1(x)\bigr\}}\bigl[py-j_m(x,y)\bigr]\,,
$$
and check that \eqref{S1d} is its maximal solution. The solution to
this variational problem is given by
$$
\mathcal H (x,p)= p\dot x(x,p)-j_m(x,\dot x(x,p)) \,,
$$
where
$$
\dot
x(x,p)=\frac12(F_1(x)+F_0(x))+\frac12(F_1(x)-F_0(x))\sqrt{\frac{\e(x,p)}{(\e(x,p)+4)}}
$$
and
$$
\epsilon(x,p)=\left(
(F_1(x)-F_0(x))\frac{p}{\sqrt{r(0,1|x)r(1,0)|x)}}+\frac{r(0,1|x)-r(1,0|x)}{\sqrt{r(1,0|x)r(0,1|x)}}\right)^2.
$$
It can be checked that the  Hamilton-Jacobi equation $\cH(x,\nabla
S)=0$ holds.
  We refer the
reader to \cite{RC} for these computations and for more details on
the one dimensional models on a bounded domain.

\bigskip

In the case of the triangular domain with unitary jump rates
discussed in subsection \ref{Ex} we have
$$
\bar{F}(x,y)=\left(\frac 13-x, \frac 13-y\right)
$$
and
\begin{equation}
\bar{F}^+(x,y)=\left(x-\frac{x^2}{x^2+y^2+(1-x-y)^2},
y-\frac{y^2}{x^2+y^2+(1-x-y)^2}\right)\,. \label{fff}
\end{equation}
The vector field $\bar{F}$ has an unique stable equilibrium point
$\left(\frac 13,\frac 13\right)\in \Omega$, which is  globally
attractive. The vector field $\bar{F}^+$ has $\left(\frac 13,\frac
13\right)$ as unique stable equilibrium point in $\O$, which is
globally attractive for all other points of $\O$. Moreover $\bar
F^+$ has also unstable equilibrium points belonging to $\partial
\Omega$. These are given by $\left(0,0\right)$, $\left(0,\frac
12\right)$, $\left(0,1\right)$, $\left(\frac 12,0\right)$,
$\left(1,0\right)$ and $\left(\frac 12,\frac 12\right)$.

The validity of the H--Theorem can be checked directly recalling the
expression \eqref{entr} and computing
$$
\nabla S(x,y)\cdot \bar{F}(x,y)=\frac 13\left(9-\frac 1x -\frac
1y-\frac{1}{1-x-y}\right)\,.
$$
The above expression is negative due to the fact  $\frac 1x+\frac
1y+\frac 1z\geq 9$ when $x$, $y$ and $z$ are constrained to satisfy
the relation $x+y+z=1$ (this follows from the convexity of $x
\rightarrow 1/x$). We can also compute
$$
\nabla S(x,y)\cdot \bar{F}^+(x,y)=\frac{1}{x^2+y^2+(1-x-y)^2}-3
$$
that is negative due to the fact that $\frac{1}{x^2+y^2+z^2}\leq 3$
when $x$, $y$ and $z$ are constrained to satisfy the relation
$x+y+z=1$ (again by convexity).

Finally, we point out that our general considerations allow  to
solve non trivial variational problems. Consider for example the
problem to determine the infimum \begin{equation}\inf_{\mathcal
A_{\left(x,y\right)}}\int_{-\infty}^0\Big(2-\sum_{(\s,\s')\in
W}\sqrt{\chi_\s(t)}\sqrt{\chi_{\s'}(t)}\Big)\, dt\,, \label{mintr}
\end{equation}
where $\cA_{(x,y)}$ denotes the family of continuous paths
$z(t):(-\infty,0]\rightarrow \bbR^2$, such that (i) $\lim_{t
\downarrow -\infty} z(t)= (1/3,1/3)$,  (ii) $z(0)=(x,y)$, (iii)
$\dot z(t)$ belongs to the convex hull $
  C\left\{F_1(z(t) ),F_2(z(t)),F_3(z(t))\right\}
$, while  $\left(\chi_1(t),\chi_2(t),\chi_3(t)\right)$ denotes the
unique probability measure such that $ \dot{z}
(t)=\sum_{\s=1}^3\chi_\s(t)F_\s(z(t)) $. This last identity  is
equivalent to the system
$$
\left\{
\begin{array}{l}
\chi_1(t)=1-x(t)-y(t)-\dot x(t)-\dot y(t)\,,\\
\chi_2(t)=x(t)+\dot x(t)\,,\\
\chi_3(t)=y(t)+\dot y(t)\,.\\
\end{array}
\right.
$$
Since the integrand in \eqref{mintr} is the LD functional density $
j(x(t), \chi (t) )$ (cf. \eqref{density}), while $(1/3,1/3)$ is the
equilibrium point of the vector field $\bar F$, we know that the
minimal value in \eqref{mintr} is given by  the quantity $S(x,y) $
in \eqref{entr}, after suitable renormalization. More precisely, the
solution of \eqref{mintr} equals
$$
-\log x -\log y -\log(1-x-y) -3\log 3\,.
$$
Moreover, the minimizer in \eqref{mintr} is obtained by
time--inversion of the solution of the Chauchy problem
$$
\begin{cases}
(\dot x(t),\dot y(t) )=\bar F^+(x(t),y(t))\,,\\
(x(0),y(0))=(x,y)\,,
\end{cases}
$$
where the vector field $\bar F^+$ is computed in \eqref{fff}.

\subsubsection{Generalizations of identity \eqref{kkk} } As already
observed, the validity of the FD relation for PDMPs with invariant
measures of the form \eqref{guess}  follows from the identity
\eqref{kkk}. This key identity can be rewritten as
\begin{equation}
j_m(x,\dot x)=\nabla W\cdot \dot x +\mathcal G(x,-\dot x)\,,
\label{ver}
\end{equation}
where $W=S$ and $\cG (x, -\dot x)= j_m^+ (x, -\dot x)$. In this
subsection, we desire to present a conjecture related to
\eqref{ver}. In general, the computation of the quasi--potential
\eqref{quasipot} for PDMPs not having invariant measure of the form
\eqref{guess} is non trivial. We will discuss a specific example of
this type in the next subsection. A possible purely variational
approach to this problem is as follows. Suppose we can decompose the
dynamic LD rate density as in \eqref{ver}, where now  $\mathcal
G(x,\dot x)$ is a nonnegative function, which is  zero only when
$\dot x= G(x)$, with $G$ a vector field having $x_*$ has the unique
global attractive equilibrium point. Then from the general arguments
in subsection \ref{degregori} we have that $W$ coincides in fact
with the quasi--potential $Q$. Inspired by the structure of the rate
functionals for PDMPs we can search for $\mathcal G$ having a
specific form. For simplicity we discuss only the case of computable
rates of the form \eqref{densitymsimplex} when the chemical part of
the generator is reversible for any $x$ and consequently $j$ is
given by \eqref{density}. In this case given a positive $\psi(\s,x)$
we can search for a $\mathcal G$ of the form
$$
\mathcal G(x,\dot x)=\sum_\s
\widetilde{\gamma}(\s|x)\chi^{-F}_\s(\dot x)-\sum_{(\s,\s')\in
W}\sqrt{\frac{\mu(\s|x)}{\mu(\s'|x)}}r(\s,\s'|x)\sqrt{\chi^{-F}_\s(\dot
 x)}\sqrt{\chi^{-F}_{\s'}(\dot
 x)}\,,
$$
where
$$
\widetilde{\gamma}(\s|x):=\sum_{\s'}r(\s',\s|x)\frac{\psi(\s',x)}{\psi(\s,x)}\,.
$$
To verify \eqref{ver} we need to find a function $W$ such that for
any $x$, $\dot x$ it holds
\begin{equation}
\sum_\s\left(\gamma
(\s|x)-\widetilde{\gamma}(\s|x)\right)\chi_\s^F(\dot x)=\nabla
W\cdot \dot x=\sum_\s\left(\nabla W\cdot
F_\s(x)\right)\chi_\s^F(\dot x)\,. \label{inft}
\end{equation}
To derive the above condition we used $\chi_\s^F(\dot
x)=\chi_\s^{-F}(-\dot x)$. Condition \eqref{inft} is verified if and
only if for any $\s \in \Gamma$ and for any $x\in \Omega$ we have
that $\gamma (\s|x)-\widetilde{\gamma}(\s|x)$ is the directional
derivative of $W$ at $x$ along the direction $F_\s(x)$. In this case
the vector field $G$ is given by
$$
G(x)=-\frac{1}{Z(x)}\sum_\s\frac{\psi^2(\s,x)}{\mu(\s|x)}F_\s(x)\,,
$$
where $Z(x):=\sum_\s\frac{\psi^2(\s,x)}{\mu(\s|x)}$. If we can find
the positive functions $\psi$ in such a way that the above
requirements are satisfied then the function $W$ obtained in
\eqref{inft}, appropriately normalized, coincides with the
quasi--potential.  In this case we obtain also that $\bar F^+$ in
fact coincides with $G$. We do not discuss this issue here.

\subsubsection{PDMPs on the 1D torus} We now consider  the    PDMPs
with $\Omega=\mathbb R/\mathbb Z$ discussed in Section \ref{torus}.
If the equilibrium condition \eqref{eqtorus} holds,  then we know
that the invariant measure has the form \eqref{guess} and therefore
the validity of relation \eqref{quintettsim} follows from the above
discussion. We consider here the general case, without assuming
\eqref{eqtorus}.

Starting from the exact expression \eqref{solper} of the invariant
measure we can derive the LD  functional of $\hat \rho_\l$. We have
that
\begin{multline}
\lim_{\l \to \infty} \hat \rho _\l (x)= \lim_{\lambda \to
\infty}\lambda^{-1}\log
\Big(\rho_\lambda(x,0)+\rho_\lambda(x,1)\Big)=\\\sup_{y\in[x,x+1]}\Big(S(y)-S(x)\Big)+c=:-W(x)\,,
\label{rateinvp}
\end{multline}
where $c$ is an appropriate additive constant related to the
normalization factor $k=k(\l)$ in \eqref{solper}. Formula
\eqref{rateinvp} follows from the fact that for arbitrary
$a(\lambda)$ and $b(\lambda)$ it holds
$$
\lim_{\lambda \to +\infty}\lambda^{-1}\log
\bigl(a(\lambda)+b(\lambda)\bigr)=\max\bigl\{\lim_{\lambda \to
+\infty}\lambda^{-1}\log a(\lambda),\lim_{\lambda \to
+\infty}\lambda^{-1}\log b(\lambda)\bigr\}\,,
$$
and from the Laplace theorem \cite{DZ}. Note that the function $W$
defined in \eqref{rateinvp}, due to the validity of \eqref{salt},
satisfy the periodicity condition $W(x)=W(x+1)$ and consequently it
can be  interpreted as a function on the torus $\O=\mathbb R/\mathbb
Z$. The constant $c$ appearing in \eqref{rateinvp} can be computed
observing that, since $\int _{\O} \hat \rho _\l (x)dx =1$, the
Laplace theorem implies that $ \inf_{x\in[0,1]}W(x)=0$. Therefore,
it must be
$$
c=-\sup_{x\in[0,1]}\sup_{y\in[x,x+1]}\Big(S(y)-S(x)\Big)\,.
$$

\smallskip

The above function $W$ is the LD functional for the measure $\hat
\rho _\l $. If  the function $S$ is periodic, then it is simple to
check that $W(x)= S(x)-\min _{y \in [0,1]} S(y)$. In the general
case, $W$ is a nonnegative function that can be flat on subregions
of $\O$.

\smallskip

Let us now verify the validity of the FD relation
\eqref{quintettsim}, with $\cL (x,\dot x)= j_m (x,\dot x)$, $\cL^+
(x, \dot x)= j_m^+(x, \dot x)$ and  $V=W$.  For simplicity, we
assume the same conditions discussed after \eqref{ris2}. The
function $y(x)$ is defined as in the paragraph  below \eqref{ris2}.
First we need to compute $\nabla W$. We have that in the points
where $W$  is differentiable it holds
\begin{equation}
\nabla W(x)=\nabla S(x)-\nabla S(y(x)) \nabla y(x)\,. \label{diffW}
\end{equation}
When $y(x)\in (x,x+1)$ we have that \eqref{diffW} becomes
\begin{equation}
\nabla W(x)=\nabla S(x)\,,
\end{equation}
due to the fact that $\nabla S(y(x))=0$. When $y(x)\in
\left\{x,x+1\right\}$ then \eqref{diffW} becomes
\begin{equation}
\nabla W(x)=0\,,
\end{equation}
due to the fact that $\nabla S(x)=\nabla S(x+1)$ and $\nabla y
(x)=1$. Note that this second alternative holds for any $x$ if the
vector fields $F_0$ and $F_1$ have the same sign so that in this
case $W$ is identically zero.

The lagrangian $j_m(x,\dot{x})$ can be computed by means of
\eqref{due} and \eqref{densitymsimplex}, getting that its expression
coincides with \eqref{quellagia}. The lagrangian $j_m^+(x,\dot {x})$
can be computed by  adapting the arguments in \cite{FGRmat}. We get
that its expression can be obtained from \eqref{sissi}  where
instead of the $\l$--dependent rates $r^+$ we have to use their
asymptotic limit value given by \eqref{genlamlmd}. We obtain for all
$x \in \O$ and $\dot x \in \cC (-F_0(x),-F_1 (x) )$
\begin{eqnarray}
j^+_m(x,\dot{x})=\left(\sqrt{\frac{r(1,0|x)B(x)(F_1(x)+\dot{x})}{F_1(x)-F_0(x)}}
-\sqrt{\frac{r(0,1|x)(\dot{x}+F_0(x))}{B(x)(F_0(x)-F_1(x))}}\right)^2\,,
\end{eqnarray}
where  $B(x):=\frac{F_0(x)}{F_1(x)} \frac{r(0,1|y(x))F_1(y(x))}
{r(1,0|y(x))F_0(y(x))}$.

We can now compute $j_m(x,\dot{x})-j_m^+(x,-\dot{x})$ obtaining
\begin{eqnarray}
& &
\dot{x}\frac{\Big[r(1,0|x)\Big(1+B(x)\Big)-r(0,1|x)\Big(1+B^{-1}(x)\Big)\Big]}{F_1(x)-F_0(x)}+\nonumber
\\
& &
\frac{r(0,1|x)\Big(F_1(x)+F_0(x)B^{-1}(x)\Big)-r(1,0|x)\Big(F_0(x)+F_1(x)B(x)\Big)}{F_1(x)-F_0(x)}\,.
\label{labello}
\end{eqnarray}
The validity of the FD relation for points $x$ where $W$ is
differentiable follows now directly from the fact that  if $y(x)\in
\left\{x,x+1\right\}$ then $B(x)=\frac{r(0,1|x)}{r(1,0|x)}$, while
if  $y(x)\in (x,x+1)$ then $B(x)=-\frac{F_0(x)}{F_1(x)}$ (recall
\ref{bacetto}). In particular the second term in \eqref{labello} is
identically zero.

Finally we point out that  in the general case  the quasi--potential
cannot be defined directly as in \eqref{quasipot}. Indeed,  the
hypothesis of existence, uniqueness   and global attractiveness of
the equilibrium point of $\bar F$ could be violated. We cannot then
identify directly $W$ with the quasi--potential.

\section{A Gallavotti--Cohen--type symmetry}
\label{GC} In this section we briefly discuss a
Gallavotti-Cohen--type (G-C) symmetry for PDMPs. Let us briefly
recall a result of \cite{LS}.
 Consider  an
involution $\cR$ on the path space of a stochastic process, i.e. a
map from the path space into itself such that $\mathcal R^2=\1$.
Assume also that the measure $\mathbb P_{st}\circ \mathcal R^{-1}$
is absolutely continuous w.r.t. $\mathbb P_{st}$, where $\mathbb
P_{st}$ denotes the stationary measure of the process. Then the
random variable
\begin{equation}
W_T:=-\frac{1}{2T}\log\frac{d\left(\mathbb P_{st}\circ \mathcal
R^{-1}\right)}{d\mathbb P_{st}}\Big|_{t\in[-T,T]}\, \label{GCgen}
\end{equation}
satisfies the G-C--type symmetry
\begin{equation}
\mathbb E_{st}\left(e^{-sW_T}\right)= \mathbb
E_{st}\left(e^{-(1-s)W_T}\right)\,,\label{GCst}
\end{equation}
where $\mathbb E_{st}$ denotes the expectation w.r.t. $\mathbb
P_{st}$.

\smallskip

Differently from the examples discussed in \cite{LS}, for  PDMPs it
is natural to consider involutions $\cR$ different from time
reversal. Indeed, take  a trajectory
$\bigl\{x(t),\sigma(t)\bigr\}_{t\in [-T,T]}$ of the PDMP, i.e.  an
element of $C([-T,T],\Omega)\times D([-T,T],\Gamma)$ such that for
any continuity point $t\in[-T,T]$ of $\bigl\{\sigma(t)\bigr\}_{t\in
[-T,T]}$ it holds
\begin{equation}
\dot{x}(t)=F_{\sigma(t)}(x(t))\,. \label{comp}
\end{equation}
Then, the  time reversed trajectory $ \mathcal
T\bigl[\bigl\{x(t),\sigma(t)\bigr\}_{t\in
[-T,T]}\bigr]=\bigl\{x(-t),\sigma(-t)\bigr\}_{t\in [-T,T]} $ is
typically not a trajectory of the PDMP due to the fact that if the
vector fields $F_\sigma$ are not identically zero then condition
\eqref{comp} is violated.  In the case of PDMPs the absolutely
continuous condition is equivalent to the preservation of relation
\eqref{comp}. This means that for any trajectory
$\bigl\{x(t),\sigma(t)\bigr\}_{t\in [-T,T]}$ also $\mathcal R
\bigl[\bigl\{x(t),\sigma(t)\bigr\}_{t\in [-T,T]}\bigr]$ has to
satisfy condition \eqref{comp}. We need then to find an involution
$\mathcal R$ on the path space preserving  relation \eqref{comp}.

This is easily done for the following class of models. Consider a
PDMP such that for any $\sigma\in \Gamma$ and for any $x\in \Omega$
there exists a unique $\sigma'\in \Gamma$ such that
$F_{\sigma'}(x)=-F_\sigma(x)$. We call $R_{x}$ the involution on
$\Gamma$ that associates to every $\sigma$ the corresponding
$\sigma'$ characterized as above. Then the map $\cR$ defined as
\begin{equation}
\mathcal R\bigl[\bigl\{x(t),\sigma(t)\bigr\}_{t\in
[-T,T]}\bigr]:=\bigl\{x(-t),R_{x(-t)}\sigma(-t)\bigr\}_{t\in [-T,T]}
\label{invR}
\end{equation}
is an involution on the path space  preserving \eqref{comp}.  An
example of such a PDMP is given by $\Omega=\mathbb R^2/\mathbb Z^2$,
$\Gamma=\left\{1,2,3,4\right\}$ and vector fields $F_i=e_i$, where
$e_1$ and $e_2$ constitute the canonical basis of $\mathbb R^2$ and
$e_3=-e_1$, $e_4=-e_2$. Then $\cR_x 1= 3$, $\cR _x 2=4$, $\cR _x
3=1$ and $\cR_x 4=2$.

If   we consider models satisfying additional assumptions  we obtain
an explicit form of the functional \eqref{GCgen} having  a direct
physical interpretation. More precisely,  we assume that the jump
rates satisfy the  generalized detailed balance condition
\begin{equation}\label{gen_DBE}
r(\sigma,\sigma'|x) =\exp\{H(\sigma,x)-H(\sigma',x)\}
r(R_x\sigma',R_x\sigma|x)\,,
\end{equation}
for a suitable  energy function $H:\Gamma\times \Omega\to \mathbb
R$,  and that the function $\g(\cdot|\cdot)$ satisfies
\begin{equation}
\gamma(\sigma|x)=\gamma(R_x\sigma|x)\,, \ \ \ \ \ \forall
(x,\sigma)\in \O\times \G\,. \label{cond2lavendetta}
\end{equation}
 This happens for example if we define  the rates as
\begin{equation}
r(\sigma,\sigma'|x):=\exp\bigl\{[H(\sigma,x)-H(\sigma',x) ]
/2\bigr\}\,, \label{sim1}
\end{equation}
for an energy function $H$ satisfying the symmetry condition
\begin{equation}
H(\sigma,x)=H(R_x\sigma,x)\,, \ \ \ \ \ \forall (x,\sigma)\in
\Omega\times \G\,. \label{sim2}
\end{equation}

Assuming \eqref{gen_DBE} and \eqref{cond2lavendetta}, it is   easy
to compute \eqref{GCgen} using  standard methods  for jump processes
(see for example \cite{KL}). One gets up to boundary terms
\begin{equation}\label{elefante}
W_T=\frac{1}{2T}\sum_{i}\left\{H(\sigma(\tau_i^-),x(\tau_i))-H(\sigma(\tau_i),x(\tau_i))\right\}\,.
\end{equation}
The boundary terms are due to the fact that in \eqref{GCgen} we are
considering stationary measures. In the case of compact phase space
$\Omega \times \Gamma$ they are negligible in the limit of diverging
$T$. In the above formula \eqref{elefante} the sum is over the jump
times $\tau_i$ of $\left\{\sigma(t)\right\}_{t\in[-T,T]}$ and we
denote the left limit as $\sigma(t^-):=\lim_{\Delta \downarrow
0}\sigma(t-\Delta)$. Since for any trajectory it holds
\begin{multline*}
 H(\sigma(T),x(T))-H(\sigma(-T),x(-T))=\int_{-T}^T\nabla
H(\sigma(s),x(s))\cdot \dot{x}(s) \, ds  \\
+
\sum_{i}\left\{H(\sigma(\tau_i),x(\tau_i))-H(\sigma(\tau_i^-),x(\tau_i))\right\}\,,
\end{multline*}
in the case of bounded energy functions $H$  we can  derive from
\eqref{elefante} that
\begin{equation}
W_T=\frac{1}{2T}\int_{-T}^T \nabla H(\sigma(s),x(s))\cdot
\dot{x}(s)\, ds+o(1)\,,\label{ultima}
\end{equation}
which is  the {\it averaged mechanical work done on the system by
the external force fields $\nabla H$}, apart negligible errors as
$T\uparrow \infty$. Trivially, for  PDMPs, \eqref{ultima} coincides
with
\begin{equation}
W_T=\frac{1}{2T}\int_{-T}^T \nabla H(\sigma(s),x(s))\cdot
F_{\sigma(s)}(x(s)) \, ds+o(1)\,.\label{ultimavera}
\end{equation}

\appendix

\section{Extended generator}\label{ext_gen}
%When studying the stationary measure of the  PDMP   $\bigl(
%x(\cdot), \s (\cdot) \bigr)$ we will use properties of its  Markov
%generator $L$. In order to recall the  definition of $L$, we denote
%by $\bbP^\l_{x,\s} $ and $\bbE ^\l _{x,\s}$ the law of the process
%$\bigl( x(\cdot), \s (\cdot) \bigr)$ starting in $(x,\s)$ and the
%associated expectation, respectively.
We first recall the definition of the  Markov generator of the PDMP.
 A bounded measurable function
$f: \O \times \G \rightarrow \bbR$ is said to belong to the domain
$\cD (L)$ of $L$ if the functions
\begin{equation}\label{grefimefi} \O\times \G \ni (x, \s)
\rightarrow t^{-1} \left[ \bbE ^\l_{x, \s} \bigl( f (x_t, \s_t)
\bigr) - f (x,\s)\right]
\end{equation}
converge uniformly (i.e. w.r.t. the uniform norm $\|\cdot
\|_\infty$) to a bounded measurable function $g$ as $t \downarrow
0$. In this case, one sets $Lf :=g$.

As discussed in \cite{D2}, if the jump rates $r(\s,\s'|x)$ are not
uniformly bounded, it is a difficult task to characterize exactly
the domain $\cD (L)$ of the generator $L$. Moreover, $\cD (L)$ could
not contain very regular functions. Let us stress this last point by
means of a simple example discussed in more detailed in Appendix
\ref{papi}.
 We take $\O=(0,1)$, $\G=\{0, 1 \}$, $F_{0}(x) =-1$,
 $F_{1} (x) = 1$, $r(0,1|x )=1/x$, $r (1,0|x)=1/(1-x)$.
 The  associated PDMP satisfies all our
 assumptions. Indeed,  $L_c [x]$ has a unique invariant
 measure, the number of jumps in a finite interval is finite a.s. due to \eqref{sciopero} (see Appendix \ref{papi}), while
  the mechanical confinement in $\O$ is implied by
 \eqref{nonlomesso}. As discussed in  Appendix \ref{papi},  the very regular  function $f(x,\s)=\s $ does
 not belong to the domain $\cD (L)$.

On the other hand, by standard computations, it is simple to prove
that if the rates are bounded (as in the case that $\O$ is the
$d$--dimensional torus), then functions $f(x,\s)$ which are bounded
and $C^1$ in $x$ belong to the domain $\cD (L)$ of the generator and
and $Lf$ equals \eqref{gen}.
 For general rates, the same computations allow to
get the same conclusions for  functions $f(x,\s)$ that are bounded,
$C^1$ in $x$ and with compact support inside $\O$.  In order to have
a unified treatment, it is convenient to work with a weaker
definition of generator $L$ introduced by Davis (see \cite{D1},
\cite{D2}), which allows a simple characterization of the domain
$\cD(L)$ and is strong enough to develop stochastic calculus for
PDMPs.  From now on, $L$ will denote the {\it  extended generator},
whose domain $\cD (L)$ is given by the set of measurable functions
$f: \O\times \G \rightarrow \bbR$ with the following property: there
exists a measurable function $h: \O\times \G \rightarrow \bbR$ such
that the function $ t\rightarrow h(x(t), \s(t) ) $ is integrable
$\bbP^{\l}_{x,\s} $--a.s. for all $(x,\s) \in \O\times \G$ and the
process
$$ C _t ^f := f \bigl(x(t), \s(t)\bigr) - f\bigl(x(0), \s (0)\bigr) - \int _0 ^t h
\bigl(x(s), \s(s) \bigr) ds
$$
is a local martingale. Then, one sets $Lf :=h$. We have recalled
here the definition of the extended generator for completeness, the
reader non familiar with local martingales can skip it. We only use
some consequences of the definition.  In particular,  we recall that
$L$ is an extension of the classical Markov generator, the domain
$\cD (L)$ of the extended generator admits a simple characterization
and it includes all bounded functions $f(x,\s)$ which are $C^1$ in
$x$ (see Theorem (26.14) and Remark (26.16) in \cite{D2}). Moreover,
for all functions $f$ in $\cD (L)$, $Lf$ is given by \eqref{gen}. We
point out that the theory in \cite{D2} is developed under the
assumption that for any starting point $(x,\s)$ the number $N_t$ of
jumps in the interval $[0,t]$ has finite expectation. As already
observed, this condition is implied for example by
\eqref{sciopero0}.

\section{An example of 1D PDMP with singular features}\label{papi}

We consider the 1D PDMP such that  $\O=(0,1)$, $\G=\{0, 1 \}$,
$F_{0}(x) =-1$,
 $F_{1} (x) = 1$, $r(0,1|x )=1/x$, $r (1,0|x)=1/(1-x)$.
 This  PDMP satisfies all our
 assumptions. Indeed,  $L_c [x]$ has a unique invariant
 measure $\mu(\cdot |x)$ given by \eqref{rev1D}, while
  the mechanical confinement in $\O$ is implied by
 \eqref{nonlomesso}. Moreover, we claim that
the number of jumps in a finite interval is finite a.s. due to
\eqref{sciopero} (note that \eqref{sciopero0} is violated). To this
aim suppose by contradiction that  the  family of jump times $\t_k$
is a sequence converging to some $\t_*<\infty$. Before time $\t_*$
the mechanical state must be eventually in $(0,3/4]$ or in $[1/4,1)$
(otherwise it should evolve with arbitrarily large velocity). Let us
consider for example the first case. Then, the system must be
infinite times in the chemical state $\s=1$ and, once it jumps into
$\s=1$, it remains in this chemical state for a random time
typically of order one. This is in contradiction with the fact that
$\t_{k+1}-\t_k$ converges to zero.

\medskip

Let us now take $\l=1$ and  show another special feature of our
simple PDMP: the regular function $f(x,\s)=\s $ does
 not belong to the domain of the classical (i.e. not extended) Markov generator.
 To this aim, let us start in
 the point $(x_0,  0)$. Then, the r.h.s. of \eqref{grefimefi} is
 simply $ t^{-1} \bbP^1  _{x_0, 0} ( \s_t= 1)$. If \eqref{grefimefi} has
 to converge uniformly to a bounded function as $t\downarrow 0$, then it must be
 \begin{equation}\label{violazione}
 \limsup _{t \downarrow 0} \sup _{x_0 \in (0,1) }t^{-1} \bbP^1 _{x_0, 0} ( \s_t=
 1)<\infty\,.
\end{equation}
It is simple to check that the above condition is violated, thus
implying our claim. Indeed,  $\bbP _{x_0, 0}^1 ( \s_t=
 1)$ can be bounded from below by the probability that the process
 makes only one chemical jump in the time interval $[0,t]$.
 Therefore, taking $x_0<t$, we get
 \begin{multline}
\bbP^1 _{x_0, 0} ( \s_t =1) \geq \int _0 ^{x_0}   \g(0|x_0 -s) e^{-
\int _0 ^s  \g(0|x_0-u) du -\int _s ^t  \g
(1|x_0-s+v-s)dv } ds =\\
\int _0 ^{x_0}  (x_0-s)^{-1}  e^{-\int _0 ^s  (x_0-u)^{-1}  du -\int
_s ^t (1- x_0+2s -v)^{-1} dv}ds=\\
x_0^{-1} \int _0 ^{x_0}  \frac{ 1-x_0+2s-t}{1-x_0+s} ds \geq
x_0^{-1} \int _0 ^{x_0}  ( 1-x_0+2s-t) ds=1-t \,.
%\geq1-x_0
 \end{multline}
 Therefore, the supremum over $x_0 \in (0,1)$ in the l.h.s. of \eqref{violazione}
 is at least $(1-t)/t$.
 This implies that  \eqref{violazione} is violated.

\smallskip Finally, we come back to the observations about the
existence of the invariant measure collected in Section \ref{inv_e
rev}. We take $(x_0, \s_0)$ as initial state and write $\nu_t$ for
the distribution at time $t$. By compactness arguments, we know that
the sequence of probability measures $\tilde \nu_t:= t^{-1} \int_0^t
\nu _s ds $ admits a  subsequence weakly converging to a probability
measure $\nu_*$ on the closure $\bar \O \times \G$. Let us show that
$\nu_*$ has support on $\O \times \G$, thus implying that $\nu_*$
describes a steady state of the PDMP. Consider the interval
$I_\e=(0,\e)$, $\e<1$. When the mechanical state enters in the
interval $I_\e$,  the chemical state of the system  must be $0$.
After a time of order $O(\e)$ the system jumps into the chemical
state $1$ keeping this value for a time $O(1)$. During this interval
$x(t)$ moves on the right with constant velocity, spending at most
$O(\e)$ time inside $I_\e$. Hence, in a time interval of order
$O(1)$ the mechanical state is in  $I_\e$ for at most $O(\e)$ time.
This implies that $\tilde \nu_t ( I_\e \times \G) \leq c \e$, for
each $t$. It is simple to conclude that the limiting measure $\nu_*$
must give zero weight to $\{0\} \times \G$. The same conclusion
holds for the set $\{1\} \times \G$, thus proving that $\nu_*(
\{0,1\} \times \G)=0$.

\section{An example of 1D PDMP with a finite number of jumps}
\label{apeape}

We take here $\Omega=[0,1]$ and $\Gamma=\left\{0,1\right\}$. The
vector fields are given by $F_0(x)=-x$ and $F_1(x)=1-x$ and the jump
rates by $r(0,1|x)=x$ and $r(1,0|x)=1-x$.

Let us consider the process with initial condition given by
$(x^*,0)$, with $x^*$ a generic element of $\Omega$. We can easily
compute the probability that there are no chemical jumps
\begin{equation}
\mathbb P^\lambda_{(x^*,0)}\left(\sigma(t)=0, \ \forall t\in \mathbb
R^+\right)=e^{-\lambda x^*\int_0^{+\infty}e^{-t} dt}=e^{-\lambda
x^*}\geq e^{-\lambda}\,. \label{nok}
\end{equation}
A similar estimate can be obtained also if we consider the process
starting from the chemical state $1$. The above result \eqref{nok}
states that every time the process jumps into a new chemical state
$\sigma$, with positive probability uniformly bounded from below by
$e^{-\lambda}$ it will never more change its chemical state and
consequently the mechanical variable will definitely evolve
according to the ODE $\dot x=F_\sigma(x)$.  As a consequence it is
easy to derive that this PDMP has a.s. a finite number of jumps and
that the invariant measures are of the form
$$
c\delta(x)\delta_{\sigma,0}+(1-c)\delta(x-1)\delta_{\sigma,1}\,, \ \
\ \ \ c\in [0,1]\,.
$$

\section{Derivation of \eqref{cruciale} in the general
case}\label{apemaia}

We consider here the case that  there exists some $\s\in \Gamma$ for
which $\chi_\s=0$. For simplicity we assume $\chi_{\sigma_1}=0$ and
$\chi_\s>0$ for any $\s\neq \s_1$. The general case can be proved in
the same way. Let us define $ \Gamma^1:=\left\{\s\in \Gamma: \
\s\neq \s_1\right\} $ and $ W^1:=\left\{(\s,\s')\in W: \s\neq\s_1,
\s'\neq \s_1)\right\}$. We want to show that
\begin{equation}
\inf_{z\in (0,+\infty)^\Gamma}\sum_{(\s,\s')\in W}\chi_\s
r(\s,\s'|x)\frac{z_{\s'}}{z_\s}=\inf_{z\in
(0,+\infty)^{\Gamma^1}}\sum_{(\s,\s')\in W^1}\chi_\s
r(\s,\s'|x)\frac{z_{\s'}}{z_\s}\,. \label{lr}
\end{equation}
The r.h.s.  of \eqref{lr} is clearly less or equal than the l.h.s.
To prove the opposite  inequality take $z\in (0,+\infty)^{\Gamma^1}$
and consider $z^\epsilon:=(\epsilon,z)\in (0,+\infty)^{\Gamma}$. We
have that
$$
\lim_{\epsilon \to 0}\sum_{(\s,\s')\in W}\chi_\s
r(\s,\s'|x)\frac{z^\epsilon_{\s'}}{z^\epsilon_\s}=\sum_{(\s,\s')\in
W^1}\chi_\s r(\s,\s'|x)\frac{z_{\s'}}{z_\s}\,,
$$
and this implies \eqref{lr}. This argument shows that we can write
\eqref{1v} and \eqref{2v} as
\begin{align}
&j(x,\chi)=\sum_\s \chi_\s \gamma(\s|x)-\inf_{z\in
(0,+\infty)^{\Gamma^1}} \sum_{(\s,\s')\in
W^1}\chi_\s r(\s,\s'|x)\frac{z_{\s'}}{z_\s} \,,\label{1va}\\
&j^+(x,\chi)=\sum_\s \chi_\s \gamma^+(\s|x)-\inf_{z\in
(0,+\infty)^{\Gamma^1}}\sum_{(\s,\s')\in
W^1}\chi_{\s'}r(\s,\s'|x)\frac{\rho(x,\s)z_{\s}}{\rho(x,\s')z_{\s'}}
\,.\label{2va}
\end{align}
If we introduce
$\widetilde{z}_\sigma:=\frac{\chi_\s}{\rho_(x,\s)z_\s}$ for any
$\s\in \Gamma^1$ the variational expression in \eqref{2va} can be
written as
$$
\inf_{z\in (0,+\infty)^{\Gamma^1}}\sum_{(\s,\s')\in
W^1}\chi_{\s}r(\s,\s'|x)\frac{\widetilde{z}_{\s'}}{\widetilde{z}_{\s}}\,,
$$
which coincides with the variational expression in \eqref{1va}.
Relation \eqref{cruciale} now follows directly.
\bigskip

\noindent
 {\bf Acknowledgements}. This work has strongly benefited of several  discussions with  Prof. G.
 Jona--Lasinio, whom  the authors kindly thank.  They also acknowledge  Prof. E. Vanden--Eijnded
 for  useful discussions. One of the authors, D.G.,
  acknowledges the support of the G.N.F.M. Young Researcher  Project ``Statistical Mechanics of Multicomponent
Systems".

\end{document}